\documentclass[a4paper,11pt]{article}                
\usepackage{rotating,amssymb,a4wide}

\makeatletter

\@addtoreset{equation}{section}
\def\section{\@startsection {section}{1}{\z@}{-3.5ex plus -1ex minus
 -.2ex}{2.3ex plus .2ex}{\large\bf\centering}}
\def\subsection{\@startsection{subsection}{2}{\z@}{-3.25ex plus%
 -1ex minus -.2ex}{1.5ex plus .2ex}{\bf}}
\def\subsubsection{\@startsection{subsubsection}{3}{\z@}{-3.25ex plus%
 -1ex minus -.2ex}{1.5ex plus .2ex}{\sl}}
\gdef\@publabel{\hfil}
\gdef\@pubdate{\null}
\gdef\@pubnumber{\null}
\gdef\@author{\null}
\gdef\@title{\null}
\gdef\@abstract{\null}
\long\def\pubdate#1{\gdef\@pubdate{#1}}
\long\def\pubnumber#1{\gdef\@pubnumber{#1}}
\long\def\publabel#1{\gdef\@publabel{#1}}
\long\def\author#1{\gdef\@author{#1}}
\long\def\title#1{\gdef\@title{#1}}
\long\def\abstract#1{\gdef\@abstract{#1}}
\def\titlerelax{
\let\maketitle\relax
\let\settitleparameters\relax
\let\consolidatetitle\relax
\let\inittitlepage\relax
\let\finishtitlepage\relax
\let\titlepagecontents\relax
\let\multithanks\relax
\let\titlebaselines\relax
\let\@makepub\relax
\let\@maketitle\relax
\let\@makeauthor\relax
\let\@makeabstract\relax
\let\@maketitlenote\relax
\let\thanks\relax
\let\titlerelax\relax}
\def\titleclean
{\gdef\@titlenote{}
\gdef\@abstract{}
\gdef\@author{}
\gdef\@title{}
\gdef\@pubdate{}\gdef\@pubnumber{}\gdef\@publabel{}
\gdef\@dpublabel{}
}

\def\@makepub{\vbox to \z@{\hbox to \textwidth{\hfill
\@publabel \hfill
\llap{\parbox[t]{0.33\textwidth}{\raggedleft\@pubnumber}}}%
\vss}}
\def\@maketitle{\vskip 60pt \begin{center}
 {\LARGE \@title \par}
 \end{center}}

\def\@makeauthor{{%
\def\and{\smallskip {\normalsize \rm and\smallskip }}
\def\And{\medskip {\normalsize \rm and\\}\medskip}
\long\def\address##1{{\def\and{\\and\\}\medskip
                                {\small \it \\##1\\}
}}
{\centering
 \vskip 3em
 \large \lineskip .75em
 \@author}
 \par}}
\def\@makedate{\vskip 1.5em
 {\raggedright \small \noindent\@pubdate \par}}
\def\@makeabstract{\vskip 1.5em
{\small
\begin{center}
{\bf ABSTRACT\vspace{-.5em}\vspace{0pt}}
\end{center}
\quotation \@abstract \endquotation}}
\def\maketitle{\titlepage
\let\footnotesize\small \setcounter{page}{0}
\@makepub
\vfil
\@maketitle
\@makeauthor
\vfil
\@makeabstract
\@thanks
\vfil
\@makedate
\if@restonecol\twocolumn \else \eject \fi
\titlerelax \titleclean
\setcounter{footnote}{0}
}

\newcommand{\ncm}{\newcommand}

\ncm{\be}{\begin{equation}
\addtolength{\abovedisplayskip}{\extraspaces}
\addtolength{\belowdisplayskip}{\extraspaces}
\addtolength{\abovedisplayshortskip}{\extraspace}
\addtolength{\belowdisplayshortskip}{\extraspace}}
\ncm{\ee}{\end{equation}}

\ncm{\bea}{\begin{eqnarray}
\addtolength{\abovedisplayskip}{\extraspaces}
\addtolength{\belowdisplayskip}{\extraspaces}
\addtolength{\abovedisplayshortskip}{\extraspace}
\addtolength{\belowdisplayshortskip}{\extraspace}}
\ncm{\eea}{\end{eqnarray}}

\ncm{\beas}{\begin{eqnarray*}
\addtolength{\abovedisplayskip}{\extraspaces}
\addtolength{\belowdisplayskip}{\extraspaces}
\addtolength{\abovedisplayshortskip}{\extraspace}
\addtolength{\belowdisplayshortskip}{\extraspace}}
\ncm{\eeas}{\end{eqnarray*}}

\newlength{\extraspace}
\newlength{\extraspaces}
\def\all{{\hbox{\bf all}}}
\def\non{{\hbox{\bf ---}}}

\def\blank#1{}

\def\c{c} \def\c{{\hat c}}

\def\cF{{\mathcal F}}

\def\D{{\mathrm d}}
\def\dslash{{\partial}{\kern -1.3ex\hbox{\Large /}}}
\def\dz{{\D z}}

\def\h{{\hat h}}
\def\hcF{{\hat\cF}}
\def\hF{{\hat F}}
\def\hG{{\hat G}}
\def\hL{{\hat L}}
\def\hM{{\hat M}}
\def\hmu{{\hat \mu}}
\def\hphi{{\hat \phi}}
\def\hsv{{h^{sv}}} \def\hsv{{\hat h}}

\def\hwt{{\mathrm{hwt}}}

\def\I{{\mathrm i}}
\def\I{{i}}
\def\II{\ensuremath{\mathrm{II}}}
\def\III{\ensuremath{\mathrm{III}}}

\def\suh{{\widehat {su}}}

\def\tik{{\mathcal X}} 
\def\tik{\surd} 
\def\ttk{\surd\surd}

\def\tik{\times}
\def\ttk{\times\times}

\def\tik{\checkmark} 
\def\ttk{\tik\tik}

\def\vac{{\vec 0}}

\def\vec#1{{ \vert {#1} \rangle }}

\begin{document}

\def\topfraction{.8}

\pubnumber{LAVAL-PHY-97-23  \\KCL-MTH-97-41 \\June 30, 1997}
\pubdate{}

\title{
Probing integrable perturbations of conformal theories using singular 
vectors II: N=1 superconformal theories
}

\author{Pierre Mathieu
\address{D\'epartement de
Physique, Universit\'e Laval, Qu\'ebec, Canada G1K 7P4}
\medskip\And\medskip
G\'erard Watts
\address{ Department of
Mathematics, King's College London, The Strand, London, WC2R 2LS}
}

\abstract{
In this work we pursue the singular-vector analysis of the integrable
perturbations of conformal theories that was initiated in
\cite{ONE}. Here we consider the detailed study of the $N=1$
superconformal theory  and show that all integrable perturbations can
be identified from a simple singular-vector argument. We identify
these perturbations as theories based on affine Lie superalgebras and
show that the results we obtain relating two perturbations can be
understood by the extension of affine Toda duality to these theories
with fermions. We also discuss how this duality is broken in specific
cases. 
} 

\maketitle

%----------------------------------------------------------
\section{Introduction}
\label{sec.intro}

An integrable perturbation of a conformal field theory is defined to
be one which has a sufficient number of conserved quantities, which
for most cases means a single nontrivial (conformal dimension $>1$)
integral of motion. An integral of motion is conserved  in the
perturbed theory if the single-pole term in the OPE of the conserved
density with the perturbing field is a total derivative.
So far the perturbations which are known to be integrable fall (apart
{}from a few cases) into series valid for all $c$, and have been shown
to be integrable using either Zamolodchikov's counting argument
\cite{ZAMO} or Feigin and Frenkel's cohomological argument based on
affine quantum groups \cite{DUA}.

In ref \cite{ONE}, we showed that the integrable perturbations of the
conformal minimal models can also all be identified by means of a
rather simple singular-vector analysis. We take a perturbation
$\phi_{r,s}(z) {\bar\phi}_{r,s}({\bar z})$, in a $(p,p')$ minimal
model with central charge  
\be
  c(p,p') = 1 - 6 \frac{(p-p')^2}{pp'}
\;.
\ee
The holomorphic part of the perturbation has conformal dimension
$h_{r,s}$ given by:
\be
  h_{r,s}
= {(rp-sp')^2-(p-p')^2\over 4pp'}
\;\qquad  1\leq r \leq p'-1 \,,\; 1\leq s \leq p-1
\;.
\label{eq:vir}
\ee
This field has primitive singular vectors at level $rs$ and
$(p'-r)(p-s)$.  All other singular vectors are descendant of these
two.  We specialise ourself to the `space of fields' $\cF_{r,s}$, in
which the singular vector at level $rs$ of each primary field is
identically set equal to zero.  As a result, all `secondary' singular
vectors disappear, leaving only the primitive singular vector at level
$(p'-r)(p-s)$.   

The point of the singular-vector analysis of integrable perturbations
is the following: the existence of a conserved density can be easily
unravelled if, for some value of the central charge, this density is
not only a vacuum descendant but also a primary field, hence
necessarily singular.%
\footnote{That conserved densities can be
singular at particular values of $c$ was first noticed in \cite{FKM}
-- see also  \cite{EY,SV}.} 
Denote such densities as $\chi_I$ (where here
and below, $\chi_{\phi}$ denotes the degenerate field associated to
the singular vector of the module $|\phi\rangle$ in the quotient space
$\cal F$). The OPE under consideration is thus of the form 
\be
  P\times \chi_I =\chi_P
\;,
\ee
where $P$ stands for the (holomorphic part of the) perturbing field.
A total derivative pole term will automatically appear if the
difference  
\be
  \Delta \, h = h_p+h_{\chi_I}-h_{\chi_P}
\;,
\ee
(which is here the difference between the level of the vacuum and the
perturbed singular  vectors) is 2: indeed, the OPE is then
\be
  P(z) \chi_I(w)  
= {1\over (z-w)^2}[\chi_P(w) +(z-w)\, k~\partial \chi_P(w)]
\;,
\ee
where $k$ is a constant.  In the Virasoro case, explicit examination
of opes with higher order poles shows it is very unlikely that they
will give conserved quantities.\footnote{
Moreover, if there are solutions corresponding to the cases $\Delta h
>2$, these cannot be generic, i.e., they do not hold for an infinite
sequence of models (in contrast to the situation for $\Delta h = 2 $
-- see below). In that sense, they would not be viewed as a genuine
signal of integrability.}
Therefore, putting in light nontrivial (albeit singular)
conservation laws in a given perturbed theory boils down to analysing
the solutions of the condition $\Delta \, h=2$. Since the level of the
vacuum singular vector is $(p'-1)(p-1)$ and that of the perturbing
vector of type $(r,s)$ is  $(p'-r)(p-s)$, we have  
\be
  \Delta \, h
=  (p'-r)(s-1)+(p-s)(r-1)+(r-1)(s-1)=2
\;.
\ee
Having written the condition in terms of a  sum of positive terms, it
is then clear that there is a very limited number of solutions.  One
of $r$ or $s$ is necessarily 1. Setting $r=1$ yields 
\be
  (p'-1)(s-1) 
= 2\quad \Rightarrow\quad p'=2, \,s=3\quad {\rm or}\quad p'=3, \, s=2
\;.
\ee
The first case corresponds to the perturbation $\phi_{1,3}$.  When
$p'=2$, $p$ is forced to be odd, say $p=2k+1$. The dimension of the
singular conserved densities (integrals) are then $2k$
(resp. $2k-1$). The second solution establishes the integrability of
the perturbation $\phi_{1,2}$, in which case the singular conserved
densities have dimension $6k\pm2$.  It is worth emphasising that in
each case, an infinite number of conservation laws is obtained, with
dimensions varying with the central charge.  By duality \cite{DUA},
$\phi_{3,1}$ and $\phi_{5,1}$ are also integrable perturbations, and
consequently, by interchanging the role of $p$ and $p'$, $\phi_{1,5}$
is integrable.

\blank{
In addition to the Virasoro minimal models, we have also considered
$W$-type conformal theories and presented a detailed study of the
$W_3$ and the $W(2,4)$ minimal models (the latter refers to the
algebra obtained by Hamiltonian reduction of the $sp(4)$ affine
algebra).  For these cases, we also found that all integrable
perturbations can be probed by a singular-vector analysis.  There is,
however, a little complication which is that for the $W$  analogues of
the $\phi_{1,2}$ perturbation, the conserved densities can be
associated to sub-singular vectors (that is, singular only in the
quotient module $\cal F$ -- cf. ref. \cite{vddv}) instead of genuine
singular vectors.  
}

In the present paper, we extend our analysis to the minimal models of
the $N=1$ superconformal algebra, whose structure is briefly reviewed
in section \ref{sec.n=1}. So far three integrable perturbations have
already been identified: $\hG_{-1/2}{\hphi}_{1,3}$ \cite{Ma}, which
preserves the supersymmetry invariance off criticality, and
${\hphi}_{1,3}$ \cite{DeMa} and ${\hphi}_{1,5}$ \cite{Melzer} which
both break supersymmetry. These three perturbations can be identified
as affine Toda theories based on Lie superalgebras through the
free-field construction, which we present in section
\ref{sec:ffconst}. As a result, they should exhibit `duality'
inherited from the free-field construction, as noted implicitly in
\cite{Vays} and  observed in the S-matrices \cite{DGZ}. In our case
this means that we can identify each perturbation with an affine super
algebra $g$ and coupling constant $\beta$, and that the theories
$(g,\beta)$ and $(g^\vee,-1/\beta)$ (where $g^\vee$ is the dual
algebra to $g$) should be equivalent. 

The singular-vector analysis of these models is presented in
sect. \ref{sec:svanalysis}. It reveals several new features which were
not seen in \cite{ONE}: 

i) The first obvious source of novelties is related to the presence of
fermionic fields.  This leads to possible fermionic integrals of
motion. We indeed find that all singular conservation laws for the
$\hphi_{1,5}$ perturbation are fermionic. As already mentioned in the
introduction, the presence of a nontrivial conservation law is
generally taken as a reliable integrability indicator. But this is
only so for a bosonic conservation law. The situation appears to be
quite different when the nontrivial conservation law is fermionic. In
that case, even the presence of an infinite number of fermionic
conserved quantities is not sufficient for integrability.  We
illustrate this point in appendix \ref{sec:fcq} by constructing
classical evolution equations for a bosonic field interacting with a
fermionic field (not necessarily invariant under supersymmetry) that
display an infinite number of fermionic conservation laws without
being integrable.   

ii) For the $\hphi_{1,5}$ perturbation, not only are all the
predicted singular conserved quantities fermionic, but these
quantities, which are predicted at specific values of $c$, do not
extend away from these $c$-values. This is in sharp contrast with all
other cases studied previously: whenever an infinite number of
(singular) conservation laws (with dimensions varying with $c$) is
found, these always proved to be particular cases of genuine
conservation laws that exist for generic values of $c$. In subsection
\ref{ssec:wb}, we relate these virtual fermionic conservation laws to
the presence of a (severely truncated) $WB(0,m)$ symmetry at the
particular values of $c$ where they appear, the conserved quantities
being the remnants of the fermionic generator of this algebra. 

iii) The present analysis also uncovers a surprising breakdown
in duality: the singular conserved densities in the model $(g,\beta)$
are not always equal to those in $(g^\vee,-1/\beta)$. A free-field
construction explanation is worked out in sections \ref{sec:ffconst}
and \ref{sec:pcfts}. 

iv) Finally we observe systematic enhancements in the number of
conservation laws of a particular perturbation for certain values of
$c$. This is again explained in general terms from the free-field
construction in sections \ref{sec:ffconst} and \ref{sec:pcfts}, and a
more precise analysis pertaining  to certain cases is presented in
section \ref{sec:exceptio}. 

We stress that in spite of the ambiguous signals related to points (i)
and (ii), the ${\hphi}_{1,5}$ perturbation is integrable and in fact
it has exactly the same {\bf generic} conserved quantities as the
$\hphi_{1,3}$ perturbation -- up to $p,p'$ interchange -- although
none of these bosonic conserved densities ever become singular. The
duality relating the two supersymmetry-breaking perturbations is
explained in section \ref{sec:toda}. 

%-----------------------------------------------------------------
\section{$\mathbf{N=1}$ superconformal models}
\label{sec.n=1}

We start by reviewing the structure of the highest-weight
representations of the $N=1$ superconformal algebra and their singular
vectors.   
The Verma modules $\hM$
are labelled by $c$ and $h$ which
we parametrise as%
\footnote{To avoid any ambiguities with their Virasoro relatives, all
quantities which refer to the superconformal algebra  are
distinguished by a hat. Note that $\c$ should not be confused with
$3c/2$ used in the `older literature'.} 
\be
  c = \c(t) = \frac{15}2 - \frac 3t - 3t
\;,\;\;
  h = \h_{r,s}(t) = \frac{r^2 - 1}{8 t} + \frac{(s^2 - 1)t}8
          + \frac{ 1 - rs}{4}
          + \frac{ 1 - (-1)^{r+s}}{32}
\;.
\label{eq:param}
\ee
We shall usually denote a highest-weight vector (whether the
highest-weight vector of smallest $\h$ in a Fock or Verma module, or
an embedded highest-weight vector) corresponding to these values of
$\h$ and $\c$ by $\vec{r,s}$, suppressing the dependence on $t$. A
Neveu-Schwarz (Ramond) Verma module $\hM_{\c,\h}$ has a highest weight
at level $rs/2$ if $\c = \c(t)$ and $\h=\h_{r,s}(t)$ for $r,s$
positive integers with $r-s$ even (odd). For any pair of positive
integers we shall define the space $\hcF_{\c(t),\h_{r,s}(t)}$ as the
quotient of $\hM_{\c(t),\h_{r,s}(t)}$ by the Verma module
$\hM_{\c(t),\h_{r,s}(t)+rs/2}$ generated by the singular vector at
level $rs/2$. We shall usually denote $\hcF_{\c(t),\h_{r,s}(t)}$ by
$\hcF_{r,s}$ and $\hcF_{1,1}$ by $\hcF$. These spaces play an
important role in Toda theory as the spaces of fields.  

We shall take our conserved quantities to be polynomials in $\hL$ and
$\hG$, in which case they correspond to vectors in $\hcF$. 
The possible embedding patterns of singular vectors are identical to
those of the Virasoro algebra, and it is only in cases $\III_-$ and
$\III_+$ (in the notation of  \cite{FF})
that there are singular vectors in $\hcF$ which may provide singular
conserved quantities. From now on,  we shall restrict to $t >0$, in
which case the values of $\c$ for which  $\hcF$ has singular  vectors
are exactly the superconformal minimal models (including the limiting
case $\c=3/2$).

The superconformal minimal models are characterised by two positive
integers $(p,p')$ such that%
\footnote{
The fact that $p,p'$ must satisfy eq. (\ref{eq.ppmin}) can  be most 
neatly seen  from the non-unitary coset description \cite{NU}: 
$$
  {\suh(2)_m\oplus \suh(2)_2\over \suh(2)_{m+2}}
\;,\;\;
  {\rm with}\quad m= {s\over u}\qquad (s,u)=1
\;,\;\;
   m+2 = {2p'\over p-p'} 
\;,\;\; 
   p-p'= 2u
\;.
$$
The basic requirement is to have  $s$ and $u$ coprime integers 
(and $u\geq 1$), which leads simply to this result.}
\be
  {p-p'\over 2}\hbox{ and } p' \hbox{ are relatively prime integers}
\;,
\label{eq.ppmin}
\ee
with $\c$ given by (\ref{eq:param}) with  $t = p'/p$.
The minimal model primary fields ${\hphi}_{r,s}$ have conformal
dimensions  $\h_{r,s}(t)$ where $1\leq r< p'$, $1\leq s < p$. 
The Verma module associated to $\hphi_{r,s}$ has primary singular
vectors at level $rs/2$ and $(p'-r)(p-s)/2$ and an infinite number of 
secondary singular vectors. Fields with $r+s$ even are in the
Neveu-Schwarz (NS) sector, while $r+s$ odd are in the  Ramond (R)
sector.% 
\footnote{
Fields in the NS sector may be considered as superfields, that is,
composed of two Virasoro primary fields, with conformal dimension
differing by $1/2$.  In a superspace formalism, a field  $\phi$ in the
NS sector can be decomposed as $\phi = \varphi + \theta \psi$;
$\varphi$ and $\psi$ are usually referred to as the lower (dimension
$h$) and the upper (dimension $h+1/2$) components of the superfield
respectively.  
}  
Correspondingly, $\hcF_{r,s}$ has a single singular vector of weight
$\h_{r,s} + (p'-r)(p-s)/2 = \h_{r,2p-s} = \h_{2p'-r,s}$, 
and in particular, $\hcF = \hcF_{1,1}$ has a unique singular
vector at level $(p'-1)(p-1)/2 = \h_{1,2p-1} = \h_{2p'-1,1}$.

%----------------------------------------------------
\section{
The singular-vector analysis of the perturbed \\
$\mathbf{N=1}$ superconformal models
}
\label{sec:svanalysis}

For a superconformal model, the perturbation may be either 
a NS or a R field. In the first case, the perturbation may either
break or preserve supersymmetry. To cover all three possibilities we
have to allow perturbations of both forms 
\be
  \oint \frac{\dz}{2\pi\I}\, P(z)\;,\;\;\hbox{ and }
  \oint \frac{\dz}{2\pi\I}\, \hG_{-1/2} P(z)\;.
\ee
The first possibility covers the R  and the supersymmetry-breaking
NS perturbations, and the second the NS supersymmetry-preserving
perturbations. Equally well, we shall consider conserved densities
(which are now always NS fields -- being vacuum descendants) of the
two forms 
\be 
  \Phi_I  \;,\;\;\hbox{ and }\;\; \hG_{-1/2}\Phi_I
\;,
\ee
although we shall usually phrase our arguments in terms of the
corresponding states
\be 
  \chi_I = \Phi_I(0)\vac\;,\;\;\;\;
  \hG_{-1/2}\chi_I = \hG_{-1/2}\Phi_I(0)\vac
\;,
\ee
where $\chi_I$ is a NS highest-weight state. Finally, we need to
consider the operator product of the perturbing field with the
highest-weight state corresponding to the conserved density, and the
leading term in the OPE may again be a highest-weight state $\chi_P$
or a descendant $\hG_{-1/2}\chi_P$. 

The simplest way for the density to be conserved is that the leading
term in the OPE is a second-order pole.%
\footnote{
As in the Virasoro case, no generic solutions are expected when the
pole is not of second order.} 
There are then three cases to consider, corresponding to whether the
leading term is $\chi_P$ for $P$ a NS field, $\chi_P$ for $P$ a R
field, or $\hG_{-1/2} \chi_P$ for $P$ a NS field. In the first case,
the only possible field which can appear as the residue of the first
order pole is the total derivative $\hL_{-1}\chi_P$. In the second case,
there may additionally be a term $\hG_{-1}\chi_P$ and we must check that
this does not appear. Finally, in the third case, the two fields which
may appear are $\hL_{-1}\hG_{-1/2}\chi_P$ and $\hG_{-3/2}\chi_P$ and
we must again check the the second field does not appear.   

Since we take $t=p'/p$ with $p,p'$ satisfying (\ref{eq.ppmin}),
the only singular vector in $\hcF_{r,s}$ is of weight $\h_{r,2p-s}$ and
consequently, with 
$P=\hphi_{r,s}$,
we identify
$\chi_P = \vec{ r,2p-s }$, 
and
$\chi_I = \vec{1,2p-1}$, 
and the requirement that the leading pole be second order equates to a
condition on the values of $r,s,p$ and $p'$. 
To find this condition we first define
\be
  \Delta{\hsv} \equiv
  \h_P + \h_{\chi_I} - \h_{\chi_P} =
  \frac 12 \left[
  (p'-r)(s-1) + (p-s)(r-1) + (r-1)(s-1) \right]
\;.
\ee
As in the Virasoro case, the simplicity of the singular-vector
argument depends crucially upon the fact that the above expression can
be written as a sum of positive terms. We give in table
\ref{tab:1} the eight possible operator product expansions, the value 
$\Delta{\hsv}$ must take for the leading pole to be of order two,
whether $P$ may be NS or $R$ (indicated by a $\tik$) and 
whether the density is guaranteed to be conserved (indicated by
$\ttk$ )

\begin{table}[htb]
{\small
\renewcommand{\arraystretch}{1.2}
\[
\begin{array}{| r @{~~\times~~} r @{~~\rightarrow~~}r | c | l | l |}
\hline
\hbox{Pert.} & \hbox{Dens.} & \hbox{ } &
2 \Delta{\hsv} & \hbox{NS} & \hbox{R} \\
\hline
\hG_{-1/2}P  & \hG_{-1/2} \chi_I &          \chi_P & 2 &\ttk&    \\
\hG_{-1/2}P  & \hG_{-1/2} \chi_I & \hG_{-1/2} \chi_P & 3 &\tik&    \\
\hG_{-1/2}P  &          \chi_I &          \chi_P & 3 &\ttk&    \\
\hG_{-1/2}P  &          \chi_I & \hG_{-1/2} \chi_P & 4 &\tik&    \\
\hline
        P  & \hG_{-1/2} \chi_I &          \chi_P & 3 &\ttk&\tik\\
        P  & \hG_{-1/2} \chi_I & \hG_{-1/2} \chi_P & 4 &\tik~~&    \\
        P  &          \chi_I &          \chi_P & 4 &\ttk&\tik~~\\
        P  &          \chi_I & \hG_{-1/2} \chi_P & 5 &\tik&    \\
\hline

\end{array}
\]
\caption{Possible operator products}
\label{tab:1}
}
\end{table}

We now analyse each of the possible values of $\Delta{\hsv}$ in turn.

\subsection{$2 \Delta{\hsv} = 2$}

We are required to analyse 
\be
  (p'-r)(s-1) + (p-s)(r-1) + (r-1)(s-1) = 2
\;.
\ee
To illustrate the procedure, we give all possible results in table
\ref{tab:2} with $r,s,p,p'$ positive integers, and indicate whether or
not this corresponds to a minimal model representation $P$.

\begin{table}[htb]
{\small
\renewcommand{\arraystretch}{1.4}
\[
\begin{array}{|cccccl|cc|cc|c|}
\hline
(p-s)(r-1) &+& (p'-r)(s-1) &+& (r-1)(s-1) 
 &=2 &(r,s)&(p,p')&NS&R&\hbox{Min.}\\
\hline
2 && 0 && 0 && \left\{ \matrix{ (2,1) \cr (3,1) }\right.
             &         \matrix{ (3,-) \cr (2,-) }        
             &         \matrix{       \cr \tik  }        
             &         \matrix{ \tik  \cr \cr   }        
             &         \matrix{ \tik  \cr \tik  }        \\
1 && 1 && 0 && \multicolumn{2}{c|}{\hbox{ No solutions }}
             &&& \\
1 && 0 && 1 && (2,2) & (3,2) &&& \\
0 && 2 && 0 && \left\{ \matrix{ (1,2) \cr (1,3) }\right.
             &         \matrix{ (-,3) \cr (-,2) }        
             &         \matrix{       \cr \tik  }        
             &         \matrix{ \tik  \cr \cr   }        
             &         \matrix{ \tik  \cr \tik  }        \\
0 && 1 && 1 && (2,2) & (2,3) &&& \\
0 && 0 && 2 &&\left\{ \matrix{ (2,3) \cr (3,2) }\right.
             &         \matrix{ (3,2) \cr (2,3) }        
             &
             &         \matrix{ \tik  \cr \tik  }        
             &         \matrix{       \cr \cr   }        \\
\hline

\end{array}
\]
\caption{Solutions to $\Delta{\hsv}=1$}
\label{tab:2}
}
\end{table}

As a result, we have four solutions to the equation 
$2\Delta{\hsv}=2$ which have minimal values of $p$ and $p'$. Of these,
two have NS perturbations and two have R perturbations. However, if we
look back at table \ref{tab:1}, we see that  only the NS perturbations
ever have $2 \Delta{\hsv} = 2$, and so the R solutions are spurious.
Moreover, the remaining two NS solutions are related by interchange of
$p$ and $p'$; requiring $p'<p$ leaves us with a single solution. 

Furthermore, the conserved densities are guaranteed to be conserved in
this case. We have thus found our first integrable perturbation of a
superconformal theory by this method, which is 
\be
  \hG_{-1/2}P      = \hG_{-1/2}\hphi_{1,3}\;,\;\;
  \hG_{-1/2}\chi_I = \hG_{-1/2}\hphi_{3,1}\;,\;\;
  \chi_P = \hphi_{3,3}\;,\;\;
\;,
\ee
where $ p'=2 $ and consequently $p = 4 k$.
In these cases, the weight of the conserved density $\chi_I$ is
$\h_{3,1}(2/4k) + \frac 12 = 2k$ (cf. \cite{SV}). 

The detailed analysis of the next three cases --  
$\Delta{\hsv} = 3/2,\,2$ and $5/2$ -- is reported in  appendix D
(where the results are presented in the form of tables).  In the main
text, we confine ourself to the interpretation of the possible
solutions. 

\subsection{ $2 \Delta{\hsv} = 3$ }

In this case we find no NS perturbations in minimal models, and
instead two Ramond perturbations,
\bea
&&
  P              = \hphi_{1,2}\;,\;\;
  \hG_{-1/2}\chi_I = \hG_{-1/2}\hphi_{7,1}\;,\;\;
  p'=4 \;,\;\;
  p = 4k+2 \;,
\\
&&
  P              = \hphi_{1,4}\;,\;\;
  \hG_{-1/2}\chi_I = \hG_{-1/2}\hphi_{3,1}\;,\;\;
  p'= 2 \;,\;\;
  p = 4k \;.
\eea  
However, using the values of the coefficients of the terms
$\hG_{-1}\chi_P$ which appear in the residue of the operator product
expansion, and which are given in eqs \ref{eq.B},
we see that these do not give conserved quantities.

\subsection{ $2 \Delta{\hsv} = 4$ }

In this case we find both NS and R perturbations in minimal models. 
In the Neveu-Schwarz case, there are six possibilities:
\[
\begin{array}{rrrrr}
a)&
  \hG_{-1/2} P     = \hG_{-1/2}\hphi_{1,3}\;,\;\;
&          \chi_I =         \hphi_{5,1}\;,\;\;
&  p'= 3 \;,\;\;
&  p = 2k+3 \;,
\\
b)&
  \hG_{-1/2} P     = \hG_{-1/2}\hphi_{1,5}\;,\;\;
&          \chi_I =         \hphi_{3,1}\;,\;\;
&  p'= 2 \;,\;\;
&  p = 4k \;,
\\
c)&
   P              =         \hphi_{1,3}\;,\;\;
&   \hG_{-1/2}\chi_I = \hG_{-1/2}\hphi_{5,1}\;,\;\;
&  p'= 3 \;,\;\;
&  p = 2k+3 \;,
\\
d)&
   P              =         \hphi_{1,5}\;,\;\;
&   \hG_{-1/2}\chi_I = \hG_{-1/2}\hphi_{3,1}\;,\;\;
&  p'= 2 \;,\;\;
&  p = 4k \;,
\\
e)&
   P      = \hphi_{1,3}\;,\;\;
&   \chi_I = \hphi_{5,1}\;,\;\;
&  p'= 3 \;,\;\;
&  p = 2k+3 \;,
\\
f)&
   P      = \hphi_{1,5}\;,\;\;
&   \chi_I = \hphi_{3,1}\;,\;\;
&  p'= 2 \;,\;\;
&  p = 4k \;,
\\\noalign{and only one in the Ramond case,}
\\[-3mm]
g)&
  P      = \hphi_{1,2}\;,\;\;
&  \chi_I = \hphi_{9,1}\;,\;\;
&  p'= 5 \;,\;\;
&  p = 2k+5 \;.
\end{array}
\]
For the Ramond case, the calculation (\ref{eq.beta}) of the
coefficient of $\hG_{-1}\chi_P$ in the residue of the OPE shows that
the density is not conserved.

For cases a), b), c) and d), the calculation (\ref{eq.abNNN}) of the
coefficient of $\hG_{-3/2}\chi_P$ in the residue of the OPE shows that
the density is not conserved.

In cases e) and f) the densities are guaranteed to be conserved.

\subsubsection{Case (e) }

Here we have found that the perturbation by $\oint \hphi_{1,3}$ has
singular conserved densities for $\c(3/(2k+3))$. These are of weight
$2k+2$, so that we expect that there are conserved quantities for the 
$\hphi_{1,3}$ perturbation of weights $2, 4, 6, 8, \ldots$ which
become singular for these $\c$--values.

\subsubsection{Case (f) }

Here we have found that the perturbation by $\oint \hphi_{1,5}$ has
singular conserved densities for $\c(2/4k)$. These are of weight
$2k -1/2$, so that we expect that there are conserved quantities for the 
$\hphi_{1,5}$ perturbation of weights $3/2, 7/2, 11/2, 15/2, \ldots$
which become singular for these $\c$--values.

\subsection{ $2 \Delta{\hsv}=5$}

In that case, we have
\be
 (p'-r)(s-1)+(p-s)(r-1)+(r-1)(s-1)=5
\ee
Taking $p'<p$,
there are three solutions: $(r,s)=(2,2)$, $p'=3$ and $ p=5$,
$(r,s)=(1,2)$ with $p'=6$ and $(r,s)=(1,6)$ with $p'=2$. 
For the first  case, the perturbing field $\hphi_{2,2}$ is
identical to $\hphi_{1,3}$; this is a solution already found
previously, but viewed here from  a different form of the quotient
space $\cal F$ (i.e., the remaining primitive singular vector differs
in the two cases).  
The other two correspond to a Ramond perturbing field, which is
incompatible with $2 \Delta{\hsv}=5$ (cf. table \ref{tab:1}.)

\subsection{Summary}

We have found that the following perturbations have singular conserved
densities at the indicated values of $c$:

\begin{itemize}

\item[1]
The supersymmetry-preserving perturbation
$\hG_{-1/2}\hphi_{1,3}$ with singular conserved densities of weight $2k$
for $c=\c(2/4k)$.

\item[2]
The supersymmetry-breaking perturbation
$\hphi_{1,3}$ with singular conserved densities of weight $2k+2$
for $c=\c(3/(2k+3))$.

\item[3]
The supersymmetry-breaking perturbation
$\hphi_{1,5}$ with singular conserved densities of weight $2k - 1/2$
for $c=\c(2/4k)$.

\end{itemize}

\section{Direct construction of the conservation laws} 
\label{sec:direct}

We can now investigate by explicit computation whether the singular
conserved densities we have predicted exist, and whether they also
exist for generic $c$--values.
We give the results for the perturbations $\hG_{-1/2}\hphi_{1,3}$,
$\hphi_{1,3}$ and $\hphi_{1,5}$ in table
\ref{tab:allperts}.

In each case we have considered the equations for a general field in
$\hcF$ of low level to be a conserved density. In the first column
we give the values of $t=p'/p$ for which there is a conserved
quantity, and the multiplicity of these quantities in parentheses if
it is more than one, and `{\bf all}' if there is a conserved quantity
at that level for all $t$ (and hence for all $c$). (For instance, for
the $\hG_{-1/2}\hphi_{1,3}$ perturbation we see that there are two
conserved densities with $\Delta=4$ when $t=1$; since there is a
generic conserved density at that level (indicated by the `all'),
there is thus one extra conserved density at $t=1$.)  

In each case, we have checked whether the conserved density is a
highest weight, whether it is the super-descendant $\hG_{-1/2}\psi$ of
some field $\psi$ (modulo total derivatives) and whether that state
$\psi$ is itself ever a highest-weight state in $\hcF$. The number of
densities of each form is again given in parentheses. (Again
considering the $\hG_{-1/2}\hphi_{1,3}$ perturbation, we see that the
generic conserved density of weight 4 is always a super-descendant
$\hG_{-1/2}\psi$ and that $\psi$ is singular for $t=1/4,4$. In the
special case $t=1$, only one of the two conserved densities of weight
4 is a super-descendant and it is not singular.) 

%%%%%%%%%%%%%%%%%%%%%%%%%%%%%%%%%%%
% All perturbations
%%%%%%%%%%%%%%%%%%%%%%%%%%%%%%%%%%%
\def\te.{}
\def\te.{t=}
\begin{sidewaystable}[!bp] {
\small
\footnotesize
\scriptsize
\renewcommand{\arraystretch}{1.4} 
\[
\begin{array}{||c||c|c|c|c||c|c|c|c||c|c|c|c||} \hline

       & \multicolumn{4}{c||}{ \hG_{-1/2}\,\hphi_{1,3}}
       & \multicolumn{4}{c||}{ \hphi_{1,3}}
       & \multicolumn{4}{c||}{ \hphi_{1,5}}
\\   \cline{2-13} 
\Delta & t       & \hwt & \hG_{-1/2}\,\psi & \hG_{-1/2}\,\hwt 
       & t       & \hwt & \hG_{-1/2}\,\psi & \hG_{-1/2}\,\hwt 
       & t       & \hwt & \hG_{-1/2}\,\psi & \hG_{-1/2}\,\hwt 
\\ \hline

3/2    &\all\,(1)&\te.2,1/2& \non      & \non  
       & \non    &      &              &
       & \non    &      &              &

\\ \hline
2      &\all\,(1)& \non & \tik         & \te.1/2,2
       &\all\,(1)& \non & \tik         & \te.1/2,2
       &\all\,(1)& \non & \tik         & \te.1/2,2

\\ \hline
5/2    & \non    &      &              &
       & \non    &      &              &
       & \non    &      &              &

\\ \hline
3      & \non    &      &              &
       & \non    &      &              &
       & \non    &      &              &

\\ \hline
7/2    & 1\,(1)  & \non & \non         &
       & \non    &      &              &
       & 1/4\,(1)& \tik & \non         &

\\     & 2\,(1)  & \non & \non         &
       &         &      &              &
       &         &      &              &

\\ \hline
4      &\all\,(1)& \non & \tik         & \te.1/4,4
       &\all\,(1)&\te.3/5& \te.2       & \non
       &\all\,(1)&\te.5/3& \te.1/2     & \non

\\     & 1\,(2)  & \non & (1)          & \non
       &         &      &              &
       &         &      &              &

\\ \hline
9/2    & 1\,(1)  & \non & (1)          & \non
       & 1\,(1)  & \non & (1)          & \non
       & \non    &      &              &

\\ \hline
5      & \non    &      &              &
       & \non    &      &              &
       & \non    &      &              &

\\ \hline
11/2   & 1\,(2)  & \non &  \non        & 
       & \non    &      &              &
       & 1/6\,(1)& \tik & \non         &

\\     & 2\,(1)  & \non &  \non        & 
       &         &      &              &
       &         &      &              &

\\ \hline
6      &\all\,(1)& \non &  \tik        & \te.1/6,6
       &\all\,(1)&\te.3/7& \te.2       & \non
       &\all\,(1)&\te.7/3& \te.1/2     & \non

\\     & 1\,(4)  & \non &  (2)         & \non
       &3/2\,(2) & \non & \non         &
       &2/3\,(2) & \non & \non         &

\\     & 2\,(2)  & \non &  (1)         & \non
       &         &      &              &
       &         &      &              &

\\ \hline
13/2   & 1\,(2)  & \non &  (2)         & \non
       & 1\,(1)  & \non &  (1)         & \non
       & \non    &      &              &

\\     & 2\,(1)  & \non &  (1)         & \non 
       &         &      &              &
       &         &      &              &

\\ \hline
7      & 1\,(1)  & \non &  \non        & 
       & \non    &      &              &
       & \non    &      &              &

\\ \hline
15/2   & 1\,(5)  & \non &  (1)         & \non
       & 1\,(1)  & \non & \non         &
       & 1/8\,(1)& \tik & \non         &

\\     & 2\,(2)  & \non &  \non        & 
       &         &      &              &
       &         &      &              &

\\ \hline
8      &\all\,(1)& \non &  \tik        & \te.1/8,8
       &\all\,(1)& \non &  \te.2       & \non
       &\all\,(1)& \non &  \te.1/2     & \non

\\     & 1\,(7)  &      &  (4)         & \non
       & 1\,(2)  & \non & (1)          & \non
       &         &      &              &
   
\\     & 2\,(3)  & \non &  (2)         & \non
       &3/2\,(3) & \non &  \non        &
       &2/3\,(3) & \non &  \non        &

\\ \hline

17/2   & 1\,(4)  & \non &  (3)         & \non
       & 1\,(1)  & \non &  (1)         & \non
       & \non    &      &              &
   
\\     & 2\,(1)  & \non &  (1)         & \non
       &         &      &              &
       &         &      &              &

\\ \hline

9      & 1\,(4)  & \non &  (1)         & \non
       &3/2\,(1) & \non & \non         & 
       &2/3\,(1) & \non & \non         &
   
\\     & 2\,(1)  & \non & \non         & 
       &         &      &              & 
       &         &      &              &
   
\\ \hline
\end{array}
 \]
\caption{The conserved densities of the integrable perturbations 
        $\hG_{-1/2}\,\hphi_{1,3}$, $\hphi_{1,3}$ and $\hphi_{1,5}$}
\label{tab:allperts} }\end{sidewaystable}

As can be seen from the tables, in each case the singular conserved
quantities that we have predicted do exist and that they are indeed
highest weights (respectively super-partners of highest weights) for
supersymmetry-breaking (resp. supersymmetry-preserving)
perturbations. Moreover, as for all the cases considered in
\cite{ONE}, for the perturbations $\hG_{-1/2}\hphi_{1,3}$ and
$\hphi_{1,3}$, these conserved quantities also exist for {\bf generic}
values of $c$.

However, a new phenomenon is observed for the case of the
$\hphi_{1,5}$ perturbation, which is that the predicted fermionic
singular conserved quantities  do not generalise to {\bf generic} 
values of $c$. Secondly, we find that there are in fact conserved densities
for generic $c$-values of all even weights, and on closer inspection,
these are identical to the conserved densities of the 
$\hphi_{1,3}$ perturbation modulo $p,p'$ interchange. 

We now make some comments on the results of the singular-vector
analysis and the results in table \ref{tab:allperts}.

(i):
For each perturbation there are some values of $t$ at which there are
extra conservation laws  over and above those present for generic $t$,
namely at $t=1,2$ for $\hG_{-1/2}\hphi_{1,3}$, at $t=1,3/2$ for
$\hphi_{1,3}$ and at $t=2/3, 1/(2k)$ for $\hphi_{1,5}$. Each of these
cases can be understood from the free-field construction in sections
\ref{sec:ffconst}-\ref{sec:pcfts} and in certain cases from the
presence of identifiable extra symmetries, which we detail in section
\ref{sec:exceptio}. 

(ii):
As mentioned in the introduction,  on the basis of the free-field
construction and the connection to affine Toda field theory, we expect
the conservation laws to satisfy a duality relation, namely that the
conserved quantities of the perturbation $\hG_{-1/2}\hphi_{1,3}$ be
invariant under $t \to 1/t$, and that those of $\hphi_{1,3}$ and
$\hphi_{1,5}$ be swapped under $t \to 1/t$. This is indeed observed
for generic values of $t$. However, we notice some surprising failures
of duality in the singular-vector analysis and in table
\ref{tab:allperts}, namely: 

(ii)(a):
The conservation laws of weight $2k-1/2$ of $\hphi_{1,5}$ at
$t=1/(2k)$ are not present for $\hphi_{1,3}$ at $t=2k$.

(ii)(b):
The extra conservation laws at $t=1$ for $\hphi_{1,3}$ are not present
for $\hphi_{1,5}$.

(ii)(c):
The extra conservation laws at $t=2$ for $\hG_{-1/2}\hphi_{1,3}$ are
not present at the dual value of $t$, $1/2$.

In each case these are `enhancements' of the spectrum of conserved
quantities as mentioned in (i), which are not shared by the dual
model, as we explain in section \ref{sec:ffconst}.

(iii):
For all perturbations, the higher conservation laws for {\bf generic}
$t$ are bosonic.
However, for special
$t$-values there are also fermionic conservation laws, even for the
supersymmetry-breaking perturbations. In that respect, the following
two observations are worth making.

(iii)(a):
As already indicated in the introduction (cf. the analysis of appendix
\ref{sec:fcq}), fermionic conservation laws are not a sure sign of
integrability, even if there are an infinite number of them.  

(iii)(b):
The fermionic conservation laws do not necessarily anticommute amongst
themselves. Actually, the Jacobi identity forces this anticommutator
to either vanish or  be equal to a bosonic conservation law. We have
checked that the higher ($\Delta \geq 7/2$) fermionic singular
conservation laws of the $\hphi_{1,5}$ perturbation square to give the
bosonic conservation laws of appropriate dimension -- and, obviously,
the latter vanishes when all singular vectors are set to
zero.\footnote{This is reminiscent of the algebra of the fermionic
nonlocal charges for the classical supersymmetric Korteweg-de Vries
equation \cite{DarM}: a fermionic charge exists at each half-integer
degree and the Poisson bracket of any such charge with itself is equal
to a bosonic local charge.} Hence, whereas for the
$\hG_{-1/2}\hphi_{1,3}$ and $\hphi_{1,3}$ perturbations, every generic
conservation law becomes singular at a particular value of $c$, in the
$\hphi_{1,5}$ case, a generic conservation law never becomes singular
but its square root does at a particular value of $c$!

\section{Affine Toda theory, Lie superalgebras and duality}
\label{sec:toda}

The rest of the article is devoted to the analysis of the data
presented in the previous section.  This analysis will rely heavily on
the free-field construction.  Our first aim is to unravel the origin
of the generic duality which is a symmetry of 
the supersymmetry-preserving
perturbation and relates the two 
supersymmetry-breaking ones.  This is done in
two steps. In the present section, we introduce affine Toda field
theories (defined in terms of   Lie superalgebras) and show that the
three integrable perturbations can be put in a one-to-one
correspondence with the three affine Lie superalgebras having exactly
two simple roots. Duality, together with its occasional breakdown, is
then analysed in the following section.

\def\ab{{\alpha}}

The super Liouville field theory of one bosonic and one fermionic
field, with Lagrangian density
\be
  \frac{1}{2} \left( \partial_\mu\phi\partial^\mu\phi + 
                     \I \bar\psi\dslash\psi \right)
\; -{ 2\I}{\ab}\,\bar\psi\psi\,e^{-\I\phi/\ab} 
\;,\label{eq:sllag}
\ee
provides a model for the superconformal minimal models. In
particular it has $N=1$ superconformal symmetry with central charge
$c=\c(t)$ as in (\ref{eq:param}) with $t = 1/\ab^2$, and the fields  
\be
 \hphi_{1,m} = \exp\left( \I \frac{ (m-1) }{2\ab}\phi \right)
\label{eq:pfield}
\ee
transform as the appropriate Neveu-Schwarz primary fields of weight
$\h_{1,m}(t)$. 

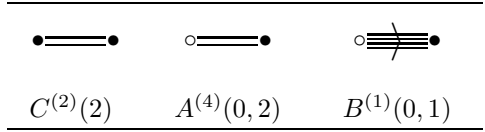
\begin{figure}[hbt]
{\renewcommand{\arraystretch}{1.2}
\footnotesize
\setlength{\unitlength}{1mm}%
\[
\begin{array}{ccc}
\hline

\begin{array}{l}{
\begin{picture}(10,10)(0,-5)
\put(0,0){\makebox(0,0)[c]{$\bullet$}}
\put(10,0){\makebox(0,0)[c]{$\bullet$}}
% two lines
\put(1,.5){\line(1, 0){8}}
\put(1,-.5){\line(1, 0){8}}
\end{picture}
}\end{array}

&

\begin{array}{l}{
\begin{picture}(10,10)(0,-5)
\put(0,0){\makebox(0,0)[c]{$\circ$}}
\put(10,0){\makebox(0,0)[c]{$\bullet$}}
% two lines
\put(1,.5){\line(1, 0){8}}
\put(1,-.5){\line(1, 0){8}}
\end{picture}
}\end{array}

&

\begin{array}{l}{
\begin{picture}(10,10)(0,-5)
\put(0,0){\makebox(0,0)[c]{$\circ$}}
\put(10,0){\makebox(0,0)[c]{$\bullet$}}
% four lines
\put(1,.9){\line(1, 0){8}}
\put(1,.3){\line(1, 0){8}}
\put(1,-.3){\line(1, 0){8}}
\put(1,-.9){\line(1, 0){8}}
% arrow
%\put(4.5,3){\line(1,-1){3}}
%\put(4.5,-3){\line(1,1){3}}
\put(5,0){\makebox(0,0)[c]{\Large$\rangle$}}
\end{picture}
}\end{array}

\\

{}~C^{(2)}(2)~~ & ~~A^{(4)}(0,2)~~ &  ~~B^{(1)}(0,1)~~

\\
\hline
\end{array}
\]

\caption{
The Dynkin diagrams of 
$C^{(2)}(2)$ 
, 
$A^{(4)}(0,2)$
and 
$B^{(1)}(0,1)$
}
\label{fig:dd}
}
\end{figure}

Olshanetsky was the first to realise that one can define integrable
field theories associated to Lie superalgebras, generalising the
bosonic affine Toda field theories to theories containing fermions
\cite{olsh1}. In particular, there are three integrable field theories
associated to Lie superalgebras which describe one bosonic and one
fermionic field, and which can be identified as perturbations of the
super-Liouville theory. They  are associated to the algebras
$B^{(1)}(0,1)$, $C^{(2)}(2)$ and $A^{(4)}(0,2)$, whose Dynkin diagrams
are given in table \ref{fig:dd}.

As explained by Liao and Mansfield \cite{LMan1}, one can simply set
the auxiliary fields to zero in the quantum theory. The argument goes
as follows.  The auxiliary fields are necessary for the classical
on-shell integrability. Since  they appear without derivatives, they
can always be eliminated. Classically, the auxiliary fields are found
(by solving the equations of motion) to be  products of
exponentials. It turns out that in the quantum theory, at imaginary
coupling, these products of exponentials vanish identically due to
normal ordering effects. Hence, they can  be ignored. The remaining
terms in the Lagrangians are the standard kinetic terms
\be
  \frac{1}{2} \left( \partial_\mu\phi\partial^\mu\phi + 
                     i \bar\psi\dslash\psi \right)
\;,
\ee
and the potentials:
\be
{\renewcommand{\arraystretch}{1}
\begin{array}{ccc}
  -{2i}{\ab}\,\bar\psi\psi\,
  \left( e^{-i\phi/\ab}  + e^{i\phi/\ab} \right)
&   \displaystyle
  -{2i}{\ab}\,\bar\psi\psi\, e^{-i\phi/\ab} 
- {\ab^2}e^{i\phi/\ab} 
&   \displaystyle
  -{2i}{\ab}\,\bar\psi\psi\, e^{-i\phi/\ab} 
- {\ab^2}e^{2i\phi/\ab} 
\\[2mm]

{}~~~~~~~~~~C^{(2)}(2)~~~~~~~~~~ &
{}~~~~~~~~~~A^{(4)}(0,2)~~~~~~~~~~ & 
{}~~~~~~~~~~B^{(1)}(0,1)~~~~~~~~~~

\end{array}}
\label{eq:pots}
\ee
As can be seen directly (i.e., by reading off the perturbation from
the second part of the potential), these three theories correspond to
the three integrable perturbations $\hG_{-1/2}\hphi_{1,3}$,
$\hphi_{1,3}$ and $\hphi_{1,5}$ that we have found. 

The conserved quantities that we are interested in are polynomials in
$\partial\phi$ and $\psi$ which commute with the terms in the
potentials (\ref{eq:pots}), whose integrals are combinations of terms
of the form:
\be
  Q^B(\beta)
= \oint \frac{ \D z}{2\pi\I} :\exp(\I \beta\sqrt 2\phi):
\;\;\;\hbox{and}\;\;\;
  Q^F(\beta)
= \oint \frac{ \D z}{2\pi\I} :\psi \exp(\I \beta \phi):
\;,
\label{eq:scs}
\ee
% with analogous expressions obtained by replacing $\beta$ by $-1/\beta$.
for various values of $\beta$. The spaces of fields polynomial in
$\partial\phi$ and $\psi$ which commute with these operators have been
studied extensively since $Q^B(\beta)$ and $Q^F(\beta)$ arise as
screening charges for the free-field constructions of the Virasoro and
$N=1$ super-Virasoro algebras respectively. We shall review in some
detail the results for the bosonic screening  charge $Q^B$ and then
present the results for $Q^F$ much more briefly; after this
digression, we will discuss the implications of this analysis for the
conserved quantities of the affine Toda theories. 

\section{The free-field construction of the Virasoro algebra\\
and the screening charge}
\label{sec:ffconst}
\label{sec:ffn=0}

Before plunging into this somewhat technical section, let us
reformulate our motivations. In the previous section, we have
introduced a correspondence between the integrable perturbations and
Toda theories.  In this section, we study the conservation laws of the
latter and their relations in theories obtained by duality with the
aim of transposing these results in the context of perturbed conformal
theories. This last point is addressed in the next section. 

Conserved integrals in the $(g,\beta)$ affine Toda theory commute with
the defining potential (given in (\ref{eq:pots2}) which is a sum of
two charges $Q$. Hence these conservations laws lie in the
intersections of the kernels of the two charges. The analysis of this
section relies on this latter point of view and the emphasis is in the
comparison of the kernel of a screening charge $Q^B(\beta)$ or
$Q^F(\beta)$ with that of the dual theory obtained by replacing
$\beta$ with $-1/\beta$.  

The presentation of this analysis is facilitated by considering first
the Virasoro case, hence by concentrating on the kernel of
$Q^B(\beta)$. With  minor modifications, these results will be used
for the description of  the bosonic part of the  
$A^{(4)}(0,2)$ and $B^{(1)}(0,1)$ potentials.

The operator $Q^B(\beta)$ has been much studied in conformal theory
because it commutes with the generator of the Virasoro algebra 
\be
  L_\beta 
= -\frac 12:(\partial\phi)^2:
  + \frac{\I}{\sqrt 2}(\beta - \frac 1\beta)\partial^2\phi
\;,
\label{eq:lb}
\ee
a property which makes it a key ingredient in the calculation of
correlation functions \cite{DF}.  A more precise formulation of this
computational approach \cite{Feld1} led to a characterisation of the
irreducible highest-weight representations of the Virasoro algebra as
cohomology spaces using maps between Fock spaces, maps which can be
considered as products of $Q^B(\beta)$. For this reason, the kernel of
$Q^B(\beta)$ has been extensively studied. 

In the following three subsections, we describe the  Virasoro
structure of the Fock spaces, the relevant maps relating the different
Fock spaces and finally the relationship between the kernels of
$Q^B(\beta)$ and $Q^B(-1/\beta)$.  These results are extended to the
supersymmetric case in the remaining subsections. 

\subsection{The free-field Fock space representations}

The modes of $L_\beta$ (\ref{eq:lb}) generate the Virasoro algebra
of central charge $c = 13 - 6 t - 6/t$ where $t = \beta^{-2}$.
They act on Fock spaces $F_\mu$ on which 
$ \alpha_0 = \oint \I \partial\phi \;\dz/(2 \pi\I) $ takes the value
$\mu$. $F_\mu$ is a highest-weight representation of the Virasoro
algebra with highest weight $\vec\mu$ whose conformal weight is
\be
  h(\mu) = \frac 12\mu^2 - \frac{\mu}{\sqrt 2}(\beta - 1/\beta)
\;.
\ee
The structure of $F_\mu$ as a representation of the Virasoro algebra
for all $\mu,\beta$ is given by Feigin and Fuchs in \cite{FF}. For our
purposes it is only necessary to consider $F_0$ and $\beta$ real. 
We can thus limit our attention to their sub-cases
$\II_-,\II_+(+),\III^{00}_-,\III^0_-(\pm)$ and $\III_-$,
 since $F_0$ with $\beta$ real must fall into one of these classes.

The classification of $F_\mu$ is given by the number of solutions of
the equation
\be
  \mu = \mu_{r,s} \equiv 
  \frac{(1-r)\beta}{\sqrt 2} - \frac{(1-s)}{\beta\sqrt 2}
\;,
\label{eq:rs}
\ee
with integral values of $r$ and $s$.
Those cases of interest to us for which (\ref{eq:rs}) gives a
single solution are as follows:

$\II_-$:
a single solution with $rs<0$.
                         
$\II_+(+)$:
a single solution with $r>0,s>0$.

For the cases $\III^*_-$ we must have $t
= p'/p$ with
$p',p$ coprime positive integers, and $\mu$ must be expressible as 
$\mu_{m+jp',m'}$ with integers $j,m,m'$,
$0\leq m<p', 0\leq m' <p$. 
Note that exactly in these cases $\mu$ does not have a unique
representation in the form (\ref{eq:rs}) but satisfies 
$\mu_{r,s} = \mu_{r+k p',s+k p}$ for any $k$.

$\III^{00}_-$:
$\mu = \mu_{jp',0}$.

$\III^{0}_-(+)$: 
(i) $ \mu = \mu_{m+jp',0}$ or 
(ii) $ \mu = \mu_{0,m+jp}$
with $j<0$.

$\III^{0}_-(-)$:
(i) $ \mu = \mu_{m+jp',0}$ or (ii) $ \mu = \mu_{0,m+jp}$
with 
$j\geq0$.

$\III_-$:
$\mu = \mu_{m + j p', m'}$ where $0 < m < p'$ and $0<m'< p$.

\begin{figure}[hbt]
{\renewcommand{\arraystretch}{1}
\footnotesize
\setlength{\unitlength}{0.4mm}%
\[
\begin{array}{ll}
\hline

\II_- &

\begin{array}{l}{
\begin{picture}(20,20)(5,-20)
\put(10,-10){\makebox(0,0)[c]{$\bullet$}}
\put(10,-15){\makebox(0,0)[c]{$u_1$}}
\end{picture}
}\end{array}

\qquad\qquad\qquad

\II_+(+)

\qquad

\begin{array}{l}{
\begin{picture}(40,20)(5,-20)
\put(28,-10){\vector(-1, 0){16}}
\put(10,-10){\makebox(0,0)[c]{$\bullet$}}
\put(30,-10){\makebox(0,0)[c]{$\circ$}}
\put(30,-15){\makebox(0,0)[c]{$v_{1}$}}
\put(10,-15){\makebox(0,0)[c]{$u_1$}}
\end{picture}
}\end{array}

\\

\hline

\III^{00}_- &

\begin{array}{l}{
\begin{picture}(136,20)(5,-20)
\put(10,-10){\makebox(0,0)[c]{$\bullet$}}
\put(30,-10){\makebox(0,0)[c]{$\bullet$}}
\put(50,-10){\makebox(0,0)[c]{$\bullet$}}
\put(70,-10){\makebox(0,0)[c]{$\bullet$}}
\put(90,-10){\makebox(0,0)[c]{$\bullet$}}
\put(110,-10){\makebox(0,0)[c]{$\bullet$}}
\put(130,-10){\makebox(0,0)[c]{$\bullet$}}
\put(10,-15){\makebox(0,0)[c]{$u_1$}}
\put(30,-15){\makebox(0,0)[c]{$u_2$}}
\put(50,-15){\makebox(0,0)[c]{$u_3$}}
\put(70,-15){\makebox(0,0)[c]{$u_4$}}
\put(90,-15){\makebox(0,0)[c]{$u_5$}}
\put(110,-15){\makebox(0,0)[c]{$u_6$}}
\put(130,-15){\makebox(0,0)[c]{$u_7$}}
\put(137,-10){\makebox(0,0)[c]{$\cdots$}}
\end{picture}
}\end{array}

\\

\III^0_-(+) &

\begin{array}{l}{
\begin{picture}(136,20)(5,-20)
\put(28,-10){\vector(-1, 0){16}}
\put(68,-10){\vector(-1, 0){16}}
\put(108,-10){\vector(-1, 0){16}}
\put(32,-10){\vector( 1, 0){16}}
\put(72,-10){\vector( 1, 0){16}}
\put(112,-10){\vector( 1, 0){16}}
\put(10,-10){\makebox(0,0)[c]{$\bullet$}}
\put(50,-10){\makebox(0,0)[c]{$\bullet$}}
\put(90,-10){\makebox(0,0)[c]{$\bullet$}}
\put(130,-10){\makebox(0,0)[c]{$\bullet$}}
\put(30,-10){\makebox(0,0)[c]{$\circ$}}
\put(70,-10){\makebox(0,0)[c]{$\circ$}}
\put(110,-10){\makebox(0,0)[c]{$\circ$}}
\put(30,-15){\makebox(0,0)[c]{$v_{1}$}}
\put(70,-15){\makebox(0,0)[c]{$v_{2}$}}
\put(110,-15){\makebox(0,0)[c]{$v_{3}$}}
\put(10,-15){\makebox(0,0)[c]{$u_1$}}
\put(50,-15){\makebox(0,0)[c]{$u_2$}}
\put(90,-15){\makebox(0,0)[c]{$u_3$}}
\put(130,-15){\makebox(0,0)[c]{$u_3$}}
\put(137,-10){\makebox(0,0)[c]{$\cdots$}}
\end{picture}
}\end{array}
\\

\III^0_-(-) &

\begin{array}{l}{
\begin{picture}(136,20)(-15,-20)
\put(28,-10){\vector(-1, 0){16}}
\put(68,-10){\vector(-1, 0){16}}
\put(108,-10){\vector(-1, 0){16}}
\put(-8,-10){\vector( 1, 0){16}}
\put(32,-10){\vector( 1, 0){16}}
\put(72,-10){\vector( 1, 0){16}}
\put(10,-10){\makebox(0,0)[c]{$\bullet$}}
\put(50,-10){\makebox(0,0)[c]{$\bullet$}}
\put(90,-10){\makebox(0,0)[c]{$\bullet$}}
\put(-10,-10){\makebox(0,0)[c]{$\odot$}}
\put(30,-10){\makebox(0,0)[c]{$\circ$}}
\put(70,-10){\makebox(0,0)[c]{$\circ$}}
\put(110,-10){\makebox(0,0)[c]{$\circ$}}
\put(-10,-15){\makebox(0,0)[c]{$v_{0}$}}
\put(30,-15){\makebox(0,0)[c]{$v_{1}$}}
\put(70,-15){\makebox(0,0)[c]{$v_{2}$}}
\put(110,-15){\makebox(0,0)[c]{$v_{3}$}}
\put(10,-15){\makebox(0,0)[c]{$u_1$}}
\put(50,-15){\makebox(0,0)[c]{$u_2$}}
\put(90,-15){\makebox(0,0)[c]{$u_3$}}
\put(117,-10){\makebox(0,0)[c]{$\cdots$}}
\end{picture}
}\end{array}

\\

\III_- &

\begin{array}{l}{
\begin{picture}(130,40)(-5,-20)
\put(28,8){\vector(-1,-1){16}}
\put(12,8){\vector( 1,-1){16}}
\put(48,8){\vector(-1,-1){16}}
\put(32,8){\vector( 1,-1){16}}
\put(68,8){\vector(-1,-1){16}}
\put(52,8){\vector( 1,-1){16}}
\put(88,8){\vector(-1,-1){16}}
\put(72,8){\vector( 1,-1){16}}
\put(108,8){\vector(-1,-1){16}}
\put(92,8){\vector( 1,-1){16}}
\put(28,-10){\vector(-1, 0){16}}
\put(12,10){\vector( 1, 0){16}}
\put(68,-10){\vector(-1, 0){16}}
\put(52,10){\vector( 1, 0){16}}
\put(108,-10){\vector(-1, 0){16}}
\put(92,10){\vector( 1, 0){16}}
\put(48,10){\vector(-1, 0){16}}
\put(32,-10){\vector( 1, 0){16}}
\put(88,10){\vector(-1, 0){16}}
\put(72,-10){\vector( 1, 0){16}}
\put(10,10){\makebox(0,0)[c]{$\square$}}
\put(10,-10){\makebox(0,0)[c]{$\bullet$}}
\put(50,10){\makebox(0,0)[c]{$\square$}}
\put(50,-10){\makebox(0,0)[c]{$\bullet$}}
\put(90,10){\makebox(0,0)[c]{$\square$}}
\put(90,-10){\makebox(0,0)[c]{$\bullet$}}
\put(30,10){\makebox(0,0)[c]{$\circ$}}
\put(30,-10){\makebox(0,0)[c]{$\circ$}}
\put(70,10){\makebox(0,0)[c]{$\circ$}}
\put(70,-10){\makebox(0,0)[c]{$\circ$}}
\put(110,10){\makebox(0,0)[c]{$\circ$}}
\put(110,-10){\makebox(0,0)[c]{$\circ$}}
%
%
%\put(8,8){\vector(-1,-1){6}}
%\put(2,-2){\vector( 1,-1){6}}
%\put(0,0){\makebox(0,0)[c]{$\odot$}}
%\put(0,6){\makebox(0,0)[c]{$v_0$}}
%
\put(8,8){\vector(-2,-1){12}}
\put(-4,-2){\vector( 2,-1){12}}
\put(-6,0){\makebox(0,0)[c]{$\odot$}}
\put(-6,6){\makebox(0,0)[c]{$v_0$}}
\put(10,15){\makebox(0,0)[c]{$w_0$}}
\put(50,15){\makebox(0,0)[c]{$w_1$}}
\put(90,15){\makebox(0,0)[c]{$w_2$}}
\put(30,15){\makebox(0,0)[c]{$v_1$}}
\put(70,15){\makebox(0,0)[c]{$v_2$}}
\put(110,15){\makebox(0,0)[c]{$v_3$}}
\put(30,-15){\makebox(0,0)[c]{$v_{-1}$}}
\put(70,-15){\makebox(0,0)[c]{$v_{-2}$}}
\put(110,-15){\makebox(0,0)[c]{$v_{-3}$}}
\put(10,-15){\makebox(0,0)[c]{$u_1$}}
\put(50,-15){\makebox(0,0)[c]{$u_2$}}
\put(90,-15){\makebox(0,0)[c]{$u_3$}}
\put(117,10){\makebox(0,0)[c]{$\cdots$}}
\put(117,-10){\makebox(0,0)[c]{$\cdots$}}
\end{picture}
}\end{array}

\\

\hline
\end{array}
\]
\caption{
The embedding diagrams of types $\II_-,\II_+(+)$ and $\III^*_-$
}
\label{fig:embed}
}
\end{figure}
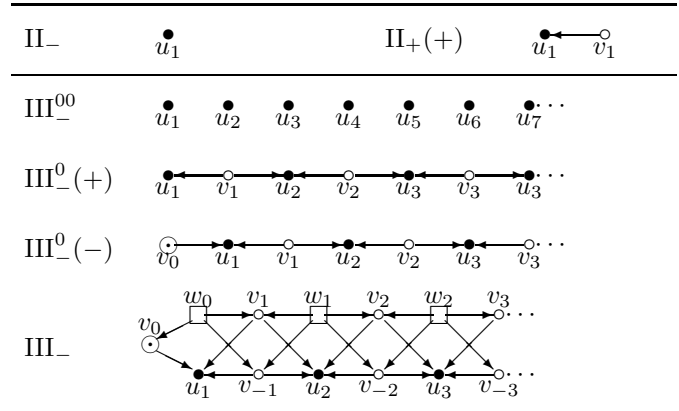

The description of the representations $F_\mu$ in \cite{FF} is in
terms of embedding diagrams; the cases of interest are presented in
table \ref{fig:embed}. In these tables, the vertices $\bullet$
correspond to highest weights of the Virasoro algebra; the vertices
$\circ$ correspond to highest weights in the quotient of $F_\mu$ by
the module generated by the vertices $\bullet$; finally, the vertices
$\square$ correspond to highest weights in the quotient of this latter
module by the submodule generated by the vertices $\circ$. The special
vertex $\odot$ corresponds to a vector of type $\circ$ which is also a
highest-weight vector itself. Finally, an arrow pointing from $a$ to
$b$ indicates that $b$ is in the submodule generated by $a$.

The dimensions of the vertices, taken from \cite{BMP} (with
$p\leftrightarrow p' $), are: 

\be
{
\begin{array}{llll}

\II_- & 
h(u_1) = h_{r,s} \\[1mm]

\II_+(+) & 
h(u_1) = h_{r,s} & h(v_1) = h_{-r,s} \\[1mm]

\III^{00}_- & 
h(u_i) = h_{(|j|+2i)p',0} 
\\[3mm]

\III^0_-(+) & 
h(u_i) = \cases{ h_{m+(j-2i+2)p',0} & \cr
                 h_{0, m'+(j-2i+2)p} &        }
& 
h(v_i) = \cases{ h_{m-(j-2i+2)p',0} & \cr
                 h_{0, m'-(j-2i+2)p} &        }
\\[4mm]

\III^0_-(-) & 
h(u_i) = \cases{ h_{m-(j+2i)p',0} & \cr
                 h_{0, m'-(j+2i)p} &        }
& 
h(v_i) = \cases{ h_{m+(j+2i)p',0} & \cr
                 h_{0, m'+(j+2i)p} &        }
\\[4mm]

\III_- & 
\matrix{ h(u_i) =  h_{-m + (|j|+2i)p',m'} & \cr
        h(w_i) =  h_{-m - (|j|+2i)p',m'} & } 
& 
h(v_i) = \cases{ h_{m+(|j|+2i)p',m'} & $i>0$ \cr
                 h_{m-(|j|-2i)p',m'} & $i<0$       }
\end{array}
}
\label{eq:hvals}
\ee

\subsection{Maps between the Fock spaces} 

As already indicated, $Q^B(\beta)=\,:\exp(\I\beta\sqrt 2\phi):$, being 
a primary field of weight 1 with respect to the Virasoro algebra
(\ref{eq:lb}), commutes with the modes of $L_\beta(z)$. As a result it
is possible to define  different maps $Q_+^{(k)}$ and $\widehat
Q_+^{(k)}$ between the various spaces $F_\mu$ from appropriate
$k$--folded products of $Q^B(\beta)$. (Such products require a
prescription for the multiple contour integrations and this rule
distinguishes  $Q_+^{(k)}$ from $\widehat Q_+^{(k)}$; the one used in
the definition of $\widehat Q_+^{(k)}$ insures that it acts
nontrivially between some spaces of interest at the boundary of the
Kac table  -- see \cite{BMP} for details). These operators have the
following properties:  

\begin{itemize}

\item[a)]
$ Q_+^{(1)} = \widehat Q_+^{(1)} = Q^B(\beta)$.
We shall also denote this by $Q_+$ for short.

\item[b)]
$Q_+^{(r)} \, Q_+^{(s)} = Q_+^{(r+s)}$, when 
$0 \leq r,s, r+s \leq p'$.

\item[c)]
$Q_+^{(p')} = 0$ for $p'> 1$.

\item[d)]
$\widehat Q_+^{(p')} \neq 0$ for $p'\geq 1$.

\end{itemize}

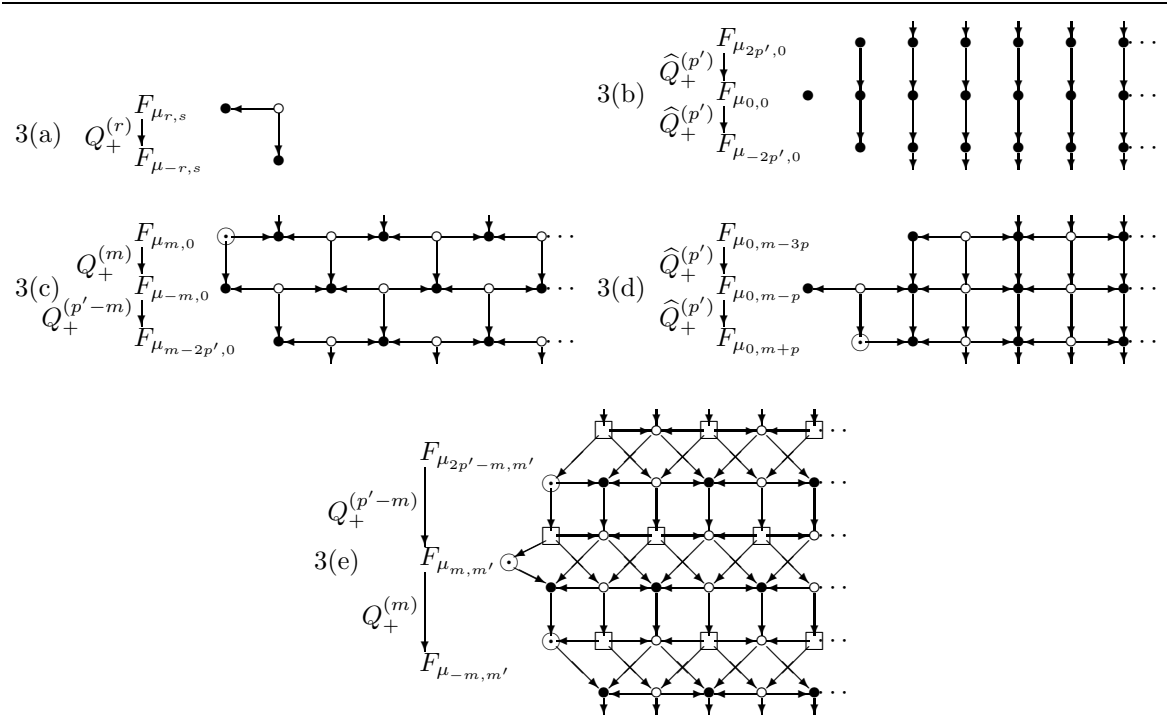
\begin{figure}[!hbt]
{\small\footnotesize
\renewcommand{\arraystretch}{1}
\setlength{\unitlength}{0.35mm}%
\[
\begin{array}{ll}
\hline

{
\begin{picture}(105,40)(-70,-20)

\put(-70,0){\makebox(0,0)[l]{\ref{fig:maps}(a)}}

\put(28,10){\vector(-1, 0){16}}
\put(10,10){\makebox(0,0)[c]{$\bullet$}}
\put(30,10){\makebox(0,0)[c]{$\circ$}}
\put(30,-10){\makebox(0,0)[c]{$\bullet$}}
\put(30,8){\vector(0,-1){16}}
\put(-25,10){\makebox(0,0)[l]{$F_{\mu_{r,s}}$}}
\put(-25,-10){\makebox(0,0)[l]{$F_{\mu_{-r,s}}$}}
%
%\put(-45,10){\makebox(0,0)[l]{$\II_+(+)$}}
%\put(-45,-10){\makebox(0,0)[l]{$\II_-$}}
\put(-22,6){\vector(0,-1){10}}
\put(-25,0){\makebox(0,0)[r]{$Q_+^{(r)}$}}

\end{picture}
}

& % \\

{
\begin{picture}(211,70)(-70,-25)

\put(-70,10){\makebox(0,0)[l]{\ref{fig:maps}(b)}}

\put(30,-10){\makebox(0,0)[c]{$\bullet$}}
\put(50,-10){\makebox(0,0)[c]{$\bullet$}}
\put(70,-10){\makebox(0,0)[c]{$\bullet$}}
\put(90,-10){\makebox(0,0)[c]{$\bullet$}}
\put(110,-10){\makebox(0,0)[c]{$\bullet$}}
\put(130,-10){\makebox(0,0)[c]{$\bullet$}}
\put(10,10){\makebox(0,0)[c]{$\bullet$}}
\put(30,10){\makebox(0,0)[c]{$\bullet$}}
\put(50,10){\makebox(0,0)[c]{$\bullet$}}
\put(70,10){\makebox(0,0)[c]{$\bullet$}}
\put(90,10){\makebox(0,0)[c]{$\bullet$}}
\put(110,10){\makebox(0,0)[c]{$\bullet$}}
\put(130,10){\makebox(0,0)[c]{$\bullet$}}
\put(30,30){\makebox(0,0)[c]{$\bullet$}}
\put(50,30){\makebox(0,0)[c]{$\bullet$}}
\put(70,30){\makebox(0,0)[c]{$\bullet$}}
\put(90,30){\makebox(0,0)[c]{$\bullet$}}
\put(110,30){\makebox(0,0)[c]{$\bullet$}}
\put(130,30){\makebox(0,0)[c]{$\bullet$}}
\put(137,-10){\makebox(0,0)[c]{$\cdots$}}
\put(137,10){\makebox(0,0)[c]{$\cdots$}}
\put(137,30){\makebox(0,0)[c]{$\cdots$}}

\put(30,8){\vector(0,-1){16}}
\put(50,8){\vector(0,-1){16}}
\put(70,8){\vector(0,-1){16}}
\put(90,8){\vector(0,-1){16}}
\put(110,8){\vector(0,-1){16}}
\put(130,8){\vector(0,-1){16}}
\put(30,28){\vector(0,-1){16}}
\put(50,28){\vector(0,-1){16}}
\put(70,28){\vector(0,-1){16}}
\put(90,28){\vector(0,-1){16}}
\put(110,28){\vector(0,-1){16}}
\put(130,28){\vector(0,-1){16}}
\put(50,-12){\vector(0,-1){6}}
\put(70,-12){\vector(0,-1){6}}
\put(90,-12){\vector(0,-1){6}}
\put(110,-12){\vector(0,-1){6}}
\put(130,-12){\vector(0,-1){6}}
\put(50,38){\vector(0,-1){6}}
\put(70,38){\vector(0,-1){6}}
\put(90,38){\vector(0,-1){6}}
\put(110,38){\vector(0,-1){6}}
\put(130,38){\vector(0,-1){6}}

\put(-25,-10){\makebox(0,0)[l]{$F_{\mu_{-2p',0}}$}}
\put(-25,10){\makebox(0,0)[l]{$F_{\mu_{0,0}}$}}
\put(-25,30){\makebox(0,0)[l]{$F_{\mu_{2p',0}}$}}

%\put(-45,-10){\makebox(0,0)[l]{$\III^{00}_-$}}
%\put(-45,10){\makebox(0,0)[l]{$\III^{00}_-$}}
%\put(-45,30){\makebox(0,0)[l]{$\III^{00}_-$}}
\put(-22,6){\vector(0,-1){10}}
\put(-22,26){\vector(0,-1){10}}
\put(-25,0){\makebox(0,0)[r]{$\widehat Q_+^{(p')}$}}
\put(-25,20){\makebox(0,0)[r]{$\widehat Q_+^{(p')}$}}

\end{picture}}

\\

{\begin{picture}(211,70)(-70,-45)

\put(-70,-10){\makebox(0,0)[l]{\ref{fig:maps}(c)}}

\put(28,-10){\vector(-1, 0){16}}
\put(68,-10){\vector(-1, 0){16}}
\put(108,-10){\vector(-1, 0){16}}
\put(32,-10){\vector( 1, 0){16}}
\put(72,-10){\vector( 1, 0){16}}
\put(112,-10){\vector( 1, 0){16}}
\put(10,-10){\makebox(0,0)[c]{$\bullet$}}
\put(50,-10){\makebox(0,0)[c]{$\bullet$}}
\put(90,-10){\makebox(0,0)[c]{$\bullet$}}
\put(130,-10){\makebox(0,0)[c]{$\bullet$}}
\put(30,-10){\makebox(0,0)[c]{$\circ$}}
\put(70,-10){\makebox(0,0)[c]{$\circ$}}
\put(110,-10){\makebox(0,0)[c]{$\circ$}}
\put(137,-10){\makebox(0,0)[c]{$\cdots$}}

%%% ABOVE
\put(48,10){\vector(-1, 0){16}}
\put(88,10){\vector(-1, 0){16}}
\put(128,10){\vector(-1, 0){16}}
\put(12,10){\vector( 1, 0){16}}
\put(52,10){\vector( 1, 0){16}}
\put(92,10){\vector( 1, 0){16}}
\put(30,10){\makebox(0,0)[c]{$\bullet$}}
\put(70,10){\makebox(0,0)[c]{$\bullet$}}
\put(110,10){\makebox(0,0)[c]{$\bullet$}}
\put(10,10){\makebox(0,0)[c]{$\odot$}}
\put(50,10){\makebox(0,0)[c]{$\circ$}}
\put(90,10){\makebox(0,0)[c]{$\circ$}}
\put(130,10){\makebox(0,0)[c]{$\circ$}}
\put(137,10){\makebox(0,0)[c]{$\cdots$}}
%%% BELOW
\put(48,-30){\vector(-1, 0){16}}
\put(88,-30){\vector(-1, 0){16}}
\put(128,-30){\vector(-1, 0){16}}
\put(52,-30){\vector( 1, 0){16}}
\put(92,-30){\vector( 1, 0){16}}
\put(30,-30){\makebox(0,0)[c]{$\bullet$}}
\put(70,-30){\makebox(0,0)[c]{$\bullet$}}
\put(110,-30){\makebox(0,0)[c]{$\bullet$}}
\put(50,-30){\makebox(0,0)[c]{$\circ$}}
\put(90,-30){\makebox(0,0)[c]{$\circ$}}
\put(130,-30){\makebox(0,0)[c]{$\circ$}}
\put(137,-30){\makebox(0,0)[c]{$\cdots$}}
%%%
\put(10,8){\vector(0,-1){16}}
\put(50,8){\vector(0,-1){16}}
\put(90,8){\vector(0,-1){16}}
\put(130,8){\vector(0,-1){16}}
\put(30,-12){\vector(0,-1){16}}
\put(70,-12){\vector(0,-1){16}}
\put(110,-12){\vector(0,-1){16}}
\put(50,-32){\vector(0,-1){6}}
\put(90,-32){\vector(0,-1){6}}
\put(130,-32){\vector(0,-1){6}}
\put(30,18){\vector(0,-1){6}}
\put(70,18){\vector(0,-1){6}}
\put(110,18){\vector(0,-1){6}}

\put(-25,10){\makebox(0,0)[l]{$F_{\mu_{m,0}}$}}
\put(-25,-10){\makebox(0,0)[l]{$F_{\mu_{-m,0}}$}}
\put(-25,-30){\makebox(0,0)[l]{$F_{\mu_{m-2p',0}}$}}

%\put(-45,10){\makebox(0,0)[l]{$\III^{0}_-(-)$}}
%\put(-45,-10){\makebox(0,0)[l]{$\III^{0}_-(+)$}}
%\put(-45,-30){\makebox(0,0)[l]{$\III^{0}_-(+)$}}
\put(-22,6){\vector(0,-1){10}}
\put(-22,-14){\vector(0,-1){10}}
\put(-25,-20){\makebox(0,0)[r]{$ Q_+^{(p'-m)}$}}
\put(-25,0){\makebox(0,0)[r]{$ Q_+^{(m)}$}}

\end{picture}}

&% \\

{\begin{picture}(211,70)(-70,-25)
\put(-70,10){\makebox(0,0)[l]{\ref{fig:maps}(d)}}
\put(28,10){\vector(-1, 0){16}}
\put(68,10){\vector(-1, 0){16}}
\put(108,10){\vector(-1, 0){16}}
\put(32,10){\vector( 1, 0){16}}
\put(72,10){\vector( 1, 0){16}}
\put(112,10){\vector( 1, 0){16}}
\put(10,10){\makebox(0,0)[c]{$\bullet$}}
\put(50,10){\makebox(0,0)[c]{$\bullet$}}
\put(90,10){\makebox(0,0)[c]{$\bullet$}}
\put(130,10){\makebox(0,0)[c]{$\bullet$}}
\put(30,10){\makebox(0,0)[c]{$\circ$}}
\put(70,10){\makebox(0,0)[c]{$\circ$}}
\put(110,10){\makebox(0,0)[c]{$\circ$}}
\put(137,10){\makebox(0,0)[c]{$\cdots$}}
%
% BELOW
\put(68,-10){\vector(-1, 0){16}}
\put(108,-10){\vector(-1, 0){16}}
\put(32,-10){\vector( 1, 0){16}}
\put(72,-10){\vector( 1, 0){16}}
\put(112,-10){\vector( 1, 0){16}}
\put(50,-10){\makebox(0,0)[c]{$\bullet$}}
\put(90,-10){\makebox(0,0)[c]{$\bullet$}}
\put(130,-10){\makebox(0,0)[c]{$\bullet$}}
\put(30,-10){\makebox(0,0)[c]{$\odot$}}
\put(70,-10){\makebox(0,0)[c]{$\circ$}}
\put(110,-10){\makebox(0,0)[c]{$\circ$}}
\put(137,-10){\makebox(0,0)[c]{$\cdots$}}
%
%ABOVE
\put(68,30){\vector(-1, 0){16}}
\put(108,30){\vector(-1, 0){16}}
\put(72,30){\vector( 1, 0){16}}
\put(112,30){\vector( 1, 0){16}}
\put(50,30){\makebox(0,0)[c]{$\bullet$}}
\put(90,30){\makebox(0,0)[c]{$\bullet$}}
\put(130,30){\makebox(0,0)[c]{$\bullet$}}
\put(70,30){\makebox(0,0)[c]{$\circ$}}
\put(110,30){\makebox(0,0)[c]{$\circ$}}
\put(137,30){\makebox(0,0)[c]{$\cdots$}}
\put(30,8){\vector(0,-1){16}}
\put(50,8){\vector(0,-1){16}}
\put(70,8){\vector(0,-1){16}}
\put(90,8){\vector(0,-1){16}}
\put(110,8){\vector(0,-1){16}}
\put(130,8){\vector(0,-1){16}}
\put(50,28){\vector(0,-1){16}}
\put(70,28){\vector(0,-1){16}}
\put(90,28){\vector(0,-1){16}}
\put(110,28){\vector(0,-1){16}}
\put(130,28){\vector(0,-1){16}}
\put(70,-12){\vector(0,-1){6}}
\put(90,-12){\vector(0,-1){6}}
\put(110,-12){\vector(0,-1){6}}
\put(130,-12){\vector(0,-1){6}}
\put(90,38){\vector(0,-1){6}}
\put(110,38){\vector(0,-1){6}}
\put(130,38){\vector(0,-1){6}}
\put(-25,-10){\makebox(0,0)[l]{$F_{\mu_{0,m+p}}$}}
\put(-25,10){\makebox(0,0)[l]{$F_{\mu_{0,m-p}}$}}
\put(-25,30){\makebox(0,0)[l]{$F_{\mu_{0,m-3p}}$}}
%
%\put(-45,-10){\makebox(0,0)[l]{$\III^{0}_-(-)$}}
%\put(-45,10){\makebox(0,0)[l]{$\III^{0}_-(+)$}}
%\put(-45,30){\makebox(0,0)[l]{$\III^{0}_-(+)$}}
\put(-22,6){\vector(0,-1){10}}
\put(-22,26){\vector(0,-1){10}}
\put(-25,0){\makebox(0,0)[r]{$ \widehat Q_+^{(p')}$}}
\put(-25,20){\makebox(0,0)[r]{$\widehat  Q_+^{(p')}$}}
\end{picture}}

\\

\multicolumn{2}{c}
{\begin{picture}(205,130)(-80,-65)
\put(-80,0){\makebox(0,0)[l]{\ref{fig:maps}(e)}}
\put(28,8){\vector(-1,-1){16}}
\put(12,8){\vector( 1,-1){16}}
\put(48,8){\vector(-1,-1){16}}
\put(32,8){\vector( 1,-1){16}}
\put(68,8){\vector(-1,-1){16}}
\put(52,8){\vector( 1,-1){16}}
\put(88,8){\vector(-1,-1){16}}
\put(72,8){\vector( 1,-1){16}}
\put(108,8){\vector(-1,-1){16}}
\put(92,8){\vector( 1,-1){16}}
\put(28,-10){\vector(-1, 0){16}}
\put(12,10){\vector( 1, 0){16}}
\put(68,-10){\vector(-1, 0){16}}
\put(52,10){\vector( 1, 0){16}}
\put(108,-10){\vector(-1, 0){16}}
\put(92,10){\vector( 1, 0){16}}
\put(48,10){\vector(-1, 0){16}}
\put(32,-10){\vector( 1, 0){16}}
\put(88,10){\vector(-1, 0){16}}
\put(72,-10){\vector( 1, 0){16}}
\put(10,10){\makebox(0,0)[c]{$\square$}}
\put(10,-10){\makebox(0,0)[c]{$\bullet$}}
\put(50,10){\makebox(0,0)[c]{$\square$}}
\put(50,-10){\makebox(0,0)[c]{$\bullet$}}
\put(90,10){\makebox(0,0)[c]{$\square$}}
\put(90,-10){\makebox(0,0)[c]{$\bullet$}}
\put(30,10){\makebox(0,0)[c]{$\circ$}}
\put(30,-10){\makebox(0,0)[c]{$\circ$}}
\put(70,10){\makebox(0,0)[c]{$\circ$}}
\put(70,-10){\makebox(0,0)[c]{$\circ$}}
\put(110,10){\makebox(0,0)[c]{$\circ$}}
\put(110,-10){\makebox(0,0)[c]{$\circ$}}
%
%
%
%\put(8,8){\vector(-1,-1){6}}
%\put(2,-2){\vector( 1,-1){6}}
%\put(0,0){\makebox(0,0)[c]{$\odot$}}
%
\put(8,8){\vector(-2,-1){12}}
\put(-4,-2){\vector( 2,-1){12}}
\put(-6,0){\makebox(0,0)[c]{$\odot$}}
\put(117,10){\makebox(0,0)[c]{$\cdots$}}
\put(117,-10){\makebox(0,0)[c]{$\cdots$}}
%
%%% BELOW
\put(12,-32){\vector( 1,-1){16}}
\put(48,-32){\vector(-1,-1){16}}
\put(32,-32){\vector( 1,-1){16}}
\put(68,-32){\vector(-1,-1){16}}
\put(52,-32){\vector( 1,-1){16}}
\put(88,-32){\vector(-1,-1){16}}
\put(72,-32){\vector( 1,-1){16}}
\put(108,-32){\vector(-1,-1){16}}
\put(92,-32){\vector( 1,-1){16}}
\put(48,-50){\vector(-1, 0){16}}
\put(32,-30){\vector( 1, 0){16}}
\put(88,-50){\vector(-1, 0){16}}
\put(72,-30){\vector( 1, 0){16}}
\put(28,-30){\vector(-1, 0){16}}
\put(68,-30){\vector(-1, 0){16}}
\put(52,-50){\vector( 1, 0){16}}
\put(108,-30){\vector(-1, 0){16}}
\put(92,-50){\vector( 1, 0){16}}
\put(30,-30){\makebox(0,0)[c]{$\square$}}
\put(30,-50){\makebox(0,0)[c]{$\bullet$}}
\put(70,-30){\makebox(0,0)[c]{$\square$}}
\put(70,-50){\makebox(0,0)[c]{$\bullet$}}
\put(110,-30){\makebox(0,0)[c]{$\square$}}
\put(110,-50){\makebox(0,0)[c]{$\bullet$}}
\put(10,-30){\makebox(0,0)[c]{$\odot$}}
\put(50,-30){\makebox(0,0)[c]{$\circ$}}
\put(50,-50){\makebox(0,0)[c]{$\circ$}}
\put(90,-30){\makebox(0,0)[c]{$\circ$}}
\put(90,-50){\makebox(0,0)[c]{$\circ$}}
\put(117,-30){\makebox(0,0)[c]{$\cdots$}}
\put(117,-50){\makebox(0,0)[c]{$\cdots$}}
%
%%% Above
%
\put(28,48){\vector(-1,-1){16}}
\put(48,48){\vector(-1,-1){16}}
\put(32,48){\vector( 1,-1){16}}
\put(68,48){\vector(-1,-1){16}}
\put(52,48){\vector( 1,-1){16}}
\put(88,48){\vector(-1,-1){16}}
\put(72,48){\vector( 1,-1){16}}
\put(108,48){\vector(-1,-1){16}}
\put(92,48){\vector( 1,-1){16}}
\put(12,30){\vector( 1, 0){16}}
\put(48,30){\vector(-1, 0){16}}
\put(32,50){\vector( 1, 0){16}}
\put(88,30){\vector(-1, 0){16}}
\put(72,50){\vector( 1, 0){16}}
\put(68,50){\vector(-1, 0){16}}
\put(52,30){\vector( 1, 0){16}}
\put(108,50){\vector(-1, 0){16}}
\put(92,30){\vector( 1, 0){16}}
\put(30,50){\makebox(0,0)[c]{$\square$}}
\put(30,30){\makebox(0,0)[c]{$\bullet$}}
\put(70,50){\makebox(0,0)[c]{$\square$}}
\put(70,30){\makebox(0,0)[c]{$\bullet$}}
\put(110,50){\makebox(0,0)[c]{$\square$}}
\put(110,30){\makebox(0,0)[c]{$\bullet$}}
\put(10,30){\makebox(0,0)[c]{$\odot$}}
\put(50,50){\makebox(0,0)[c]{$\circ$}}
\put(50,30){\makebox(0,0)[c]{$\circ$}}
\put(90,50){\makebox(0,0)[c]{$\circ$}}
\put(90,30){\makebox(0,0)[c]{$\circ$}}
\put(117,50){\makebox(0,0)[c]{$\cdots$}}
\put(117,30){\makebox(0,0)[c]{$\cdots$}}
%
%% Vertical lines
%
\put(10,28){\vector(0, -1){16}}
\put(30,28){\vector(0, -1){16}}
\put(50,28){\vector(0, -1){16}}
\put(70,28){\vector(0, -1){16}}
\put(90,28){\vector(0, -1){16}}
\put(110,28){\vector(0, -1){16}}
\put(10,-12){\vector(0, -1){16}}
\put(30,-12){\vector(0, -1){16}}
\put(50,-12){\vector(0, -1){16}}
\put(70,-12){\vector(0, -1){16}}
\put(90,-12){\vector(0, -1){16}}
\put(110,-12){\vector(0, -1){16}}
%
%% Short vertical lines
%
\put(30,58){\vector(0, -1){6}}
\put(50,58){\vector(0, -1){6}}
\put(70,58){\vector(0, -1){6}}
\put(90,58){\vector(0, -1){6}}
\put(110,58){\vector(0, -1){6}}
\put(30,-52){\vector(0, -1){6}}
\put(50,-52){\vector(0, -1){6}}
\put(70,-52){\vector(0, -1){6}}
\put(90,-52){\vector(0, -1){6}}
\put(110,-52){\vector(0, -1){6}}
%
%% labels
%
\put(-40,0){\makebox(0,0)[l]{$F_{\mu_{m,m'}}$}}
\put(-40,-40){\makebox(0,0)[l]{$F_{\mu_{-m,m'}}$}}
\put(-40,40){\makebox(0,0)[l]{$F_{\mu_{2p'-m,m'}}$}}
%
%\put(-55,0){\makebox(0,0)[l]{$\III_-$}}
%\put(-55,-40){\makebox(0,0)[l]{$\III_-$}}
%\put(-55,40){\makebox(0,0)[l]{$\III_-$}}
%
\put(-38,36){\vector(0,-1){30}}
\put(-38,-4){\vector(0,-1){30}}
\put(-40,20){\makebox(0,0)[r]{$  Q_+^{(p'-m)}$}}
\put(-40,-20){\makebox(0,0)[r]{$  Q_+^{(m)}$}}
\end{picture}}

\\

\hline
\end{array}
\]
\caption{
The maps between the Fock modules of types $\II_-,\II_+(+)$ and
$\III^*_-$ 
}
\label{fig:maps}
}
\end{figure}

{}From Theorem 3.1, Lemmas 4.1 and 5.1 of \cite{BMP}, the action of
the various maps $Q_+^{(m)}$, $Q_+^{(p'-m)}$ and $\widehat Q_+^{(p')}$
between the spaces defined above are found to be as in figure
\ref{fig:maps}. 

It is now straightforward to read off the kernel of 
$Q_+$ on $F_{\mu_{1,1}}$ for each value of $t>0$.

$t$ irrational:
$F_{\mu_{1,1}}$ is of type  $\II_+(+)$ and if we consider 
Fig. \ref{fig:maps}(a) with $r=s=1$, we find that the kernel of
$Q_+=Q_+^{(r)}$ on $F_0$ is given by the Virasoro submodule 
generated by the vector $u_1$.

$t=1$:
In this case $F_{\mu_{1,1}}= F_{\mu_{0,0}}$ is of type $\III^{00}_-$,
and if we consider Fig. \ref{fig:maps}(b) with $p'=1$, we find that
the kernel of $Q_+=\widehat Q_+^{(p')}$ on $F_0$ is given by the
Virasoro submodule generated by the vector $u_1$.

$t=p'$, $p'$ integer:
In this case $F_{\mu_{1,1}}= F_{\mu_{1-p',0}}$ is of type
$\III^0_-(+)$ and if we consider Fig. \ref{fig:maps}(c) with
$m=p'-1$, we find that the kernel of $Q_+= Q_+^{(p'-m)}$ on $F_0$ is
given by the Virasoro submodule generated by the vectors 
$u_1,u_2,u_3,\ldots$.

$t=1/p$, $p$ integer:
In this case $F_{\mu_{1,1}}= F_{\mu_{0,1-p}}$ is of type
$\III^0_-(+)$ and if we consider Fig. \ref{fig:maps}(d) with $p'=1$ we
find that that the kernel of $Q_+=\widehat Q_+^{(p')}$ on $F_0$ is
given by the Virasoro submodule generated by the vector $u_1$.

$t=p'/p$, $p',p$ integers greater than 1:
In this case $F_{\mu_{1,1}}$ is of type
$\III_-$ and if we consider Fig. \ref{fig:maps}(e) with $m=m'=1$, we
find that the kernel of $Q_+=Q_+^{(m)}$ on $F_0$ is given by the
Virasoro submodule generated by the vectors  
$v_0,w_0,v_1,w_1,v_2,w_2,\ldots$.

\subsection{The kernels of $Q^B(\beta)$ and $Q^B(-1/\beta)$}
\label{ssec:kers}

The first point to make is that $L_\beta = L_{-1/\beta}$
and $\mu_{1,1}=0$ is invariant under $\beta \to -1/\beta$,  so that
the kernels of $Q^B(\beta)$ and $Q^B(-1/\beta)$ are related.

It is clear that in the case $\beta^2=1/t$ irrational they are in fact
equal, being exactly the Virasoro submodule of $F_{\mu_{1,1}}$ 
generated by the highest-weight state $\vec 0$.

Similarly for $t=1$, the kernels of $Q^B(1)$ and $Q^B(-1)$ are again
equal,  being again exactly the Virasoro submodule of $F_{\mu_{1,1}}$ 
generated by the highest-weight state $\vec 0$.
However for all other rational values of $\beta^2$, one or other of
the kernels is increased as follows:

For $t=1/\beta^2=1/p$ for $p>1$ integer, we have the kernel of
$Q^B(\beta)$ again being simply the submodule generated by $u_1 = \vec
0$, whereas for the dual values $t=1/\beta^2=p'>1$, the kernel of
$Q^B(-1/\beta)$ is much larger being generated by $u_1,u_2,\ldots$, so
that  
\be
  \ker Q^B(-1/\sqrt{k}) \supset \ker Q^B(\sqrt k)
\;,\;\; k=2,3,\ldots
\;.
\label{eq:enlarge2}
\ee
Notice that the argument does not depend upon the sign of $\beta$
which means that we also have 
\be
  \ker Q^B(1/\sqrt{k}) \supset \ker Q^B(-\sqrt k)
\;,\;\; k=2,3,\ldots
\;.
\label{eq:enlarge22}
\ee

For generic $t=1/\beta^2=p'/p$ with $p',p>1$ coprime integers, the
kernel of $Q^B(\beta)$ is generated by $v_0,w_0,v_1,w_1,\ldots$.  
For the dual value of $\beta$ the kernel of $Q^B(-1/\beta)$ is
generated by a similar submodule of $F_{\mu_{1,1}}$, but with respect
to the labelling in the first case it is generated by
$v_0,w_0,v_{-1},w_1,v_{-2},w_2 \ldots$.
It is clear that the kernels are not identical but it is hard to tell
whether one is systematically larger than the other.

\subsection{The free-field construction of the $N=1$ super-Virasoro
algebra and the screening charge}
\label{ssec:ffn=1}

The free-field construction of the $N=1$ super-Virasoro algebra in
terms of one free boson $\phi$ and one free fermion $\psi$ has
also been known for a long time \cite{svir}.
The algebra  generators take the form
\be
   \hL_\beta(z) 
= -{1\over 2}[:(\partial \phi)^2: 
+ :\psi\partial \psi:]+\frac{i}{2} (\beta-\frac1\beta)\,\partial^2\phi
\;,\;\;\;\;
  \hG_\beta 
= \I\,\partial\phi\,\psi +  (\beta - \frac1\beta)\,\partial\psi
\;,
\ee
and the screening charge is $Q^F(\beta)$ as in (\ref{eq:scs}). The
Neveu-Schwarz Fock spaces $\hF_{\hmu_{r,s}}$ with 
$\hmu_{r,s} = (1/2)( (1-r)\beta - (1-s)/\beta)$ and $r+s$ even are
Neveu-Schwarz highest-weight representations of the $N=1$
super-Virasoro algebra with $c = \c(t)$ and $h=\h_{r,s}(t)$ as in
(\ref{eq:param}) with $t = 1/\beta^2$.  

The analysis of the structure of these Fock representations as
representations of the $N=1$ super-Virasoro algebra   carry through
exactly as for the Virasoro algebra.  The classification of Fock
modules is still given by the number of solutions to $\mu =
\hmu_{r,s}$, with the proviso that $r-s$ must be even. The analysis of
the kernel of $Q^F(\beta)$ is also the same since the maps between the
spaces are unchanged:  one only needs to replace $h_{r,s}$ by
$\h_{r,s}$, $M$ by $\hM$, $Q^B$ by $Q^F$ etc. wherever appropriate.

Accordingly, the classification of the vacuum Fock representation
$\hF_0$ is as follows:

$t$ irrational means that $\hF_0$ is of type $\II_+(+)$.

$t = 1$ means that $\hF_0$ is of type $\III^{00}_-$.

$t=1/(2k+1)$ or $t=(2k+1)$ for $k$ a positive integer means that
$\hF_0$ is of type $\III_-^0(+)$.

For all other rational values of $t$, including $t=2k,1/(2k)$, $\hF_0$
is of the minimal type, i.e., $\III_-$.

\subsection{The kernels of $Q^F(\beta)$ and $Q^F(-1/\beta)$}
\label{ssec:kers2}

Since $\hL_\beta = \hL_{-1/\beta}$ and $\hG_\beta = \hG_{-1/\beta}$,
we know that the kernels of $Q^F(\beta)$ and $Q^F(-1/\beta)$ are
related. As for the bosonic case, we can say that
$\ker Q^F(\beta) = \ker Q^F(-1/\beta)$ 
if $\beta^2$ is irrational or takes the value 1.
If $\beta^2 = 2k+1$ then
$\ker Q^F(\beta) \subset \ker Q^F(-1/\beta)$. In all other cases there
is no clear relationship between the two kernels.

\subsection{The conserved quantities of the affine Toda theories}
\label{sec:enlarge}

The conserved quantities of the affine Toda theories are given as the
intersections of the kernels of the potential terms. 
Ignoring the $\bar z$ dependent terms and the constant prefactors of
each term (which are irrelevant for the holomorphic conserved
quantities), we write these in the following way: 
\be
{\renewcommand{\arraystretch}{1}
\begin{array}{ccc}
  \displaystyle
  Q^F(\beta) + Q^F(-\beta) 
&   \displaystyle
 Q^F(\beta) +  Q^B(-\beta/\sqrt 2)
&   \displaystyle
 Q^F(\beta) + Q^B(-\beta\sqrt 2)

\\[2mm]

{}~~~~~~~~~C^{(2)}(2)~~~~~~~~~~ &
{}~~~~~~~~~A^{(4)}(0,2)~~~~~~~~~~ & 
{}~~~~~~~~~B^{(1)}(0,1)~~~~~~~~~~

\end{array}}
\label{eq:pots2}
\ee
{}From the discussion in sections \ref{ssec:kers} and \ref{ssec:ffn=1},
it is clear that for $\beta^2$ positive and irrational, the conserved
quantities of the theories $C^2(2)(\beta)$ and $C^2(2)(-1/\beta)$ are
equal, as are those of $A^{(4)}(0,2)(\beta)$ and $B^{(1)}(0,1)(-1/\beta)$. 
Further evidence for these relations comes from the equivalence of the
S-matrices of the corresponding theories for $\beta$ imaginary. 

For $\beta^2$ positive and rational, our analysis does not allow us to
say whether the conserved densities of a theory are equal to those of
its dual. However, there are special circumstances in which we can be
certain that the number of conserved quantities of one theory are no
smaller than the conserved quantities in the dual theory. But this is
so only for $C^{(2)}(2)$ (whose potential is solely given in terms of
$Q^F$), when $\beta^2 = 3,5,\ldots$. In that case, 
$
\ker Q^F(\pm\beta) \subset \ker Q^F(\mp 1/\beta)
,
$
so that the intersection satisfies
\be
(\ker Q^F(\beta) \cap \ker Q^F(-\beta)) \subseteq
(\ker Q^F(1/\beta) \cap \ker Q^F(-1/\beta))
\;.
\ee
For the other cases, the mixing of the $Q^F$ and $Q^B$ terms in the
Toda potential prevents such a simple analysis.

\blank{
For $A^{(4)}(0,2)$ vs. $B^{(1)}(0,1)$, if $\beta^2 = 4,6,8,\ldots,$
then 
\be
\ker Q^F(\beta) \subset \ker Q^F(-1/\beta)
\;\;\hbox{ and }\;\;
\ker Q^B(-\beta/\sqrt 2) \subset \ker Q^F(\sqrt 2/\beta)
\;,
\ee
so that the intersection satisfies
\be
(\ker Q^F(\beta) \cap \ker Q^B(-\beta/\sqrt 2)) \supseteq
(\ker Q^F(1/\beta) \cap \ker Q^F(\sqrt 2/\beta))
\;.
\ee
}

In the next section we will consider the relevance of 
this discussion of duality in affine Toda theories for the results
found in sections \ref{sec:svanalysis} and \ref{sec:direct}

\section{Conserved quantities: from affine Toda field theories to
perturbed conformal field theories}
\label{sec:pcfts}

The discussion of duality in the previous two sections applies to the
affine Toda field theories. However, the results that we would like to
understand pertain to  perturbed superconformal field theories. The
relation between the two problems is quite simple: affine Toda
conserved densities are also conserved densities in the perturbed
theories if they can be written in terms of 
the generators $\hG$ and $\hL$ of the unperturbed theory.

Actually, both conditions are related: 
the space $\hcF$  
can be characterised in terms of a cohomology involving 
precisely the same operators used to characterise the Toda conserved
densities (although at potentially different values of $\beta$,
clearly).

In the Virasoro case, the irreducible vacuum representation  may be
found as the BRST cohomology of the Fock space $\hF_{0}$
where the BRST maps are of the form $Q_+^{(k)}$ or 
$\widehat Q_+^{(k)}$, whichever is appropriate
\cite{Feld1,BMP}. Similarly, the action of a primary field is given as
the action of a suitably screened vertex operator of the form
(\ref{eq:pfield}).

We assume that these results transfer to the representations of the
super-Virasoro algebra as well.
 
Our computations of the conserved quantities of the affine Toda
theories (\ref{eq:pots}) must then be altered to calculate the
intersection of the kernels of the potentials corresponding to the
perturbations of the super-Liouville action (\ref{eq:sllag}) with the
BRST cohomology of the screening charge corresponding to the potential
in the super-Liouville action.

Although there are exact formulations of the super Virasoro module
$\hcF$ in terms of the Fock module $\hF_0$ as a cohomology space, it
is only in rare cases that we are able at present to relate this to
the kernels of the perturbing potentials. However, we can certainly be
sure that the space $\hcF$ has the same structure for a perturbation
and its dual, since the superconformal generators of the unperturbed
theory are identical in these cases. As a result, to compare the
conserved densities of the perturbed superconformal models we need
only compare the kernels of the perturbing potentials. Our results are
presented below. 

\subsection{The perturbation $\hG_{-1/2}\hphi_{1,3}$}

Viewed as a perturbation of the superconformal model with 
$t = 1/\beta^2$, $\oint \hG_{-1/2}\hphi_{1,3}$ has the form
$Q^F(1/\beta)$. Consequently, we can say that the perturbations by 
$\hG_{-1/2}\hphi_{1,3}$ with $t=t_0$ and $t=1/t_0$ have equal
conserved densities when $t_0$ is irrational or 1.

For all other cases the free-field argument does not allow us to say
whether they have the same conserved quantities or not, except 
when $\hF_0$ is of type $\III^0_-(+)$. In this case we have
$\ker Q^F(1/\beta) \supset \ker Q^F(\beta)$
for $\beta^2 = 2k+1$.
This implies that the perturbations by $\hG_{-1/2}\hphi_{1,3}$ 
with $t=3,5,\ldots$ may have more conserved densities than those with 
the dual values $t=1/3,1/5,\ldots$. However, we do not observe
such an enhancement in table \ref{tab:allperts}.

\subsection{The perturbations $\hphi_{1,3}$ and $\hphi_{1,5}$}

The perturbations $\oint \hphi_{1,3}$ and $\oint \hphi_{1,5}$ viewed
as perturbations of the superconformal model with $t=1/\beta^2$ have
the forms $Q^B(1/(\beta\sqrt 2))$ and $Q^B(\sqrt 2/\beta)$
respectively. As a result, the perturbations by $\hphi_{1,3}$ with
$t=t_0$ and by $\hphi_{1,5}$ with $t=1/t_0$ have equal conserved
quantities for $t_0/2$ irrational or 1. 

For all other cases, the free-field  argument does not allow us to say
whether these models related by duality have the same conserved
quantities or not, except when
$\hF_0$ is of type $\III^0_-(+)$, when 
$\ker Q^B(1/\sqrt k) \supset \ker Q^B(-\sqrt k)$. 

This is relevant when $Q^B(1/\sqrt k)=\oint\hphi_{1,3}$, i.e., 
$t = 2/k$ and when $Q^B(1/\sqrt k)=\oint\hphi_{1,5}$, i.e., $t=1/(2k)$.

In the first case, the $\hphi_{1,3}$ perturbations at $t=2/k$ will
certainly have no fewer conserved  densities than the perturbations by
$\hphi_{1,5}$ at the dual values $t=k/2$, and 
may have more.
We see that for the first value $t=1$ that $\hphi_{1,3}$ does have
more conserved quantities than $\hphi_{1,5}$.

Similarly in the second case, the $\hphi_{1,5}$ perturbed theories at
$t=1/2k$ may have more conserved  densities than the $\hphi_{1,3}$
perturbed theories at the dual values $t=2k$. 
This is indeed borne out, since the extra fermionic conserved
quantities of $\hphi_{1,5}$ at $t=1/(2k)$ have the same weight as
$\hG_{-1/2}$ acting on the extra vector $u_2$ in the kernel of
$Q^B(1/\sqrt k)$:
since $\hphi_{1,5}|_{t=1/2k} = Q^B(1/\sqrt k)$,
we must use the value $t=1/k$ to evaluate the weight of
$u_2$ using (\ref{eq:hvals}). 
We find $h(u_2)  = h_{0,3p-1} = h_{0,3k-1} = 2k-1$, so that 
$h(u_2) + 1/2 = 2k-1/2$ as required.

In all cases where $t$ rational, it is also possible that the number
of conserved quantities is increased over the number for generic
irrational $t$. This may happen either because the intersection of the
kernel of the perturbing potential with the polynomials in $\hG$ and
$\hL$ increases, or because the dimension of the kernel itself
jumps. This is a generic reason  for the possibility of an increase in
the number of conserved quantities, but in special cases we can
provide a more detailed analysis  which also allow us to count exactly
the number of conserved quantities. These cases are treated in the
next section.

\section{Results on certain exceptional cases}
\label{sec:exceptio}
\subsection{Relation with  $\widehat{su}(2)_2$}
\label{ssec:k=2}

There is a construction of the affine algebra $su(2)$ at level 2
($c=3/2$) in terms of one free boson and one free fermion for which
the currents $E(z)$ and $F(z)$ are given exactly by $Q^F(1)$ and
$Q^F(-1)$. Consequently, the conserved currents for $C^{(2)}(0,2)$ are
the singlets of the zero-grade $su(2)$ subalgebra,
 for which the generating function is known. 
To count the number of nontrivial conserved currents it is only
necessary to factor  out the generating function of $su(2)$ singlets
by the action of $\hL_{-1}$. If the number of nontrivial conserved
densities of the Toda theory at level $n$ is $d_n$, we then have
\bea
  \sum d_n q^n 
&=& 1 + (1-q) \left[ 
  \prod_{n=1}^\infty 
  \left( \frac{1+q^{n+1/2}}{1-q^{n+1}} \right)
  \; - 1 \; \right]
\nonumber\\
&=&  1 + {q^{{\frac{3}{2}}}} + {q^2} + {q^{{\frac{7}{2}}}}+2\,{q^4}+ 
   {q^{{\frac{9}{2}}}} + 2\,{q^{{\frac{11}{2}}}} + 4\,{q^6} + 
   2\,{q^{{\frac{13}{2}}}} + {q^7} + 5\,{q^{{\frac{15}{2}}}} 
   + \ldots 
\;.
\eea
Since in this case $\hF_0$ is of type $\III^{00}_-$, the kernel of
$Q^F(1)$ is also equal to the irreducible vacuum Virasoro
representation. Therefore, all the conserved quantities found above
are entirely expressed in terms of vacuum descendants,  that is, in
terms of $\hL(z)$ and $\hG(z)$, and the number of conserved quantities of
the $\hG_{-1/2}\hphi_{1,3}$ perturbation at $t=1,c=3/2$ is also given
by the formula above. This counting agrees exactly with the results in
table \ref{tab:allperts}.% 
\footnote{  
We stress that the above character codes the number of independent
expressions built out of the modes $\hL_{-k}$ and $\hG_{-m/2}$ at each
level, modulo the singular vector $\hL_{-1}|0\rangle$ and modulo total
derivatives (i.e., modulo the left action of $\hL_{-1}$).   
Here is another way of reaching the same conclusion. At $\beta=1$, the
free-field expressions for $\hL(z)$ and $\hG(z)$ are  simply 
\[
  \hL_{\beta=1}(z) 
= -{1\over 2}[:(\partial \phi)^2:+ :\psi\partial \psi:]
\;,\qquad 
  \hG_{\beta=1}(z) 
= i\psi \partial \phi
\;.
\] 
Both $\hL_1$ and $\hG_1$ are invariant under a simultaneous change of
sign in $\phi$ and $\psi$. Since the second term of the potential of
the $\hG_{-1/2}\hphi_{1,3}$ perturbation can be obtained from the other
(the screening charge) by this transformation, we conclude that any
even differential polynomial in the free fields (hence any 
differential polynomial in
$\hL$ and $\hG$) is automatically conserved. 
}

\subsection{Cases when $h=1$}
\label{ssec:h=1}

When the perturbing field $P(z)$ has conformal weight 1, the integral
of $P(z)$ automatically commutes with all the modes
$\hL_m$. Consequently, every polynomial in $\hL$ and its derivatives is a
conserved density, so that we have a lower bound on the number of
nontrivial conserved densities given by 
\bea
  \sum d_n q^n 
&=& 1 + (1-q) \left[ 
  \prod_{n=1}^\infty 
  \left( \frac{1}{1-q^{n+1}} \right)
  \; - 1 \; \right]
\nonumber\\
&=&   1 + {q^2} + {q^4} + 2\,{q^6} + 3\,{q^8} + {q^9}
   + \ldots 
\;.
\eea
The values of $t$ for which each perturbation has weight 1 are
$t=1$ for $\hG_{-1/2}\hphi_{1,3}$, $t=3/2$ for $\hphi_{1,3}$ and
$t=2/3$ for $\hphi_{1,5}$. In these latter two cases we see that this
exhausts the nontrivial conserved densities we have found.

\subsection{$WB(0,m)$ algebras and $\hphi_{1,5}$ perturbations}
\label{ssec:wb}

The $WB(0,m)$ algebras (or fermionic $WB_m$ algebras) are W-algebras
with bosonic fields of spins $2,4,\ldots 2m$ and one fermionic field
of spin $m+1/2$ \cite{FL,Wat}.
The $N=1$ superconformal algebra is itself the $WB(0,1)$ algebra.
It is natural to ask whether the fermionic conserved quantities of
the $\hphi_{1,5}$ perturbation of weight $2k-1/2$, which occur only
for $c=\c(2/4k)$, are related to the fermionic fields of the same
weight of the $WB(0,2k-1)$ algebra.

The values of the central charge $c$ corresponding to
the $WB(0,m)$ minimal models are given by
\be
  c_m(p,p') = (m+1/2) \left[ 1 - 2m(2m-1) \frac{(p-p')^2}{pp'} \right] 
\;,
\ee
where we require $p,p' \geq 2m-1$. 
The minimal model representations are characterised by two weights of
$B_m$; the representation is in the Neveu-Schwarz / Ramond sector
according to whether the difference of the two weights is in the root
lattice or the root lattice shifted by the spinor weight of $B_m$.
There is also a restriction on the weights which is that their levels
(as affine $B_m$ representations) are equal to $p'-2m+1$ and $p-2m+1$
respectively.

The $c$ values we are interested in are $\c(2/4k)=c_1(4k,2)$. It is a
simple check that
\be
 c_1(4k,2) = c_{2k-1}(2k,2k-1)
\;.
\ee
where we identify $c_1(4k,2)$ with $c_1(2k,1)$ in view of the
constraint $p+p'=$ even%
\footnote{This is an example of a general level-rank duality
$c_n(p+n,n) = c_p(p+n,n) $.}.
Although there are no minimal representations for this value of $c$,
we can still consider the corresponding highest weights of
$WB(0,2k-1)$.  Let the $WB(0,m)$ highest weights be labelled by two
$B_m$ weights $[\lambda,\lambda'\,]$ (finite parts of affine weights
at respective level $p'-2m+1$ and $p'-2m+1$) whose conformal
dimensions in the Neveu-Schwarz sector are given by
\be
h{[\lambda,\lambda'\,]} = 
{ [p(\lambda+\rho)-p'(\lambda'+\rho)]^2-\rho^2(p-p')^2\over 2 p p'}
\ee 
with $\rho = \sum \omega_i$ 
($ \omega_i$ being the $B_m$ fundamental weights). With $m=p'=2k-1$
and $p=2k$, we find that 
\be
\begin{array}{lll}
  h[0,0] = 0
\;,\;\;
&  h[\omega_1,0]=  3/2
\;,\;\;
\\
  h[0,\omega_1]=\h_{1,3}
\;,\;\;
&  h[0,\omega_2]=\h_{1,5}
\;,\;\;
&  h[0,2\omega_1]=\h_{1,3}+1/2
\;,\ldots
\end{array}
\ee
We can indeed see that the fermionic conserved quantities we find are
exactly related to the fermionic fields of the $WB(0,2k-1)$
algebras. The three perturbations of the $WB(0,m)$ algebra by the
fields $[0,\omega_1]$, $[0,\omega_2]$ and $[0,2\omega_1]$  
correspond to the affine Toda field theories 
$A^{(4)}(0,2m)$, $A^{(2)}(0,2m-1)$ and $B^{(1)}(0,m)$ respectively.

The Zamolodchikov counting argument shows that the perturbation by
$[0,\omega_2]$ does indeed have a fermionic conserved current of
weight $m+1/2$ and we identify this as our singular conserved
quantity. This conserved quantity has also been investigated in more
detail in \cite{fermcls}.

\section{Conclusions}
\label{sec:conc}

In the superconformal case, as for the Virasoro and $W$ minimal
models, all integrable perturbations can be unravelled by the
singular-vector argument.  In fact, the situation is somewhat better
in the superconformal case since, strictly speaking, the argument in
the $W$ case must be supplemented by duality (e.g., the Virasoro
$\phi_{1,5}$ perturbed integrals of motion are never singular).  In
the superconformal case, even though the generic $\hphi_{1,5}$
perturbed integrals  are never singular either, the integrability is
revealed through an infinite sequence of non-generic singular
fermionic conservation laws.  Hence, the fermionic degree of freedom
provides us a direct handle on the integrability of the $\hphi_{1,5}$
perturbation. Phrased differently, the presence of fermionic fields
allows for conservation laws to be related to singular vectors not
directly but through their square root, a possibility that is indeed
realized in the $\hphi_{1,5}$ case.

Notice that the
singular-vector argument treats with equal simplicity standard
perturbations as well as those not easily interpreted physically,
such as the Ramond perturbations.
\footnote{
Note that while in the Ramond case, the singular-vector
argument suggests the absence of an integrable  perturbation, this
does not constitute a proof of non-existence. However, explicit
computer searches have yield no densities (other than the trivial case
$\h=1$) conserved for generic values of $c$.} 

Finally we would like to comment on the fact that duality,  clearly
broken in the $N=1$ models,  was apparently preserved in the bosonic
models  treated in \cite{ONE}. Duality is indeed broken for the models
in \cite{ONE}, but so far we have only observed it  affine Toda
theories; once we turn to Virasoro perturbed models and the spaces
$\cF_{r,s}$, duality is apparently restored.

As an example, consider the Sine-Gordon model whose potential is
$ Q^B(\beta) + Q^B(-\beta)$. For generic $\beta$, there are conserved
densities of all positive even spins \cite{DUA}. However,  from
explicit calculations, we find that for
$ \beta^2 = 1/2 $ there appear extra conserved densities of weights
$3,5,7,\ldots$; for $\beta^2 = 1/3$ of weights $5,\ldots$; for
$\beta^2=2/3$ of weights $5,7,\ldots$.
In each of these cases, the extra conserved quantity of lowest weight
corresponds to the first additional vector in the kernel of
$Q^B(\beta)$ (i.e., $u_2$). However, the additional extra conserved
quantities are not directly related to the extra vectors identified in
section \ref{ssec:kers} (i.e., $u_{i\geq 3}$); they are merely
particular descendants of these vectors in the Fock module.
However, it is clear many of these extra conserved densities will not
survive when we consider the conserved densities in $\cF$, for the
simple reason that there are no quasi-primary states in $\cF$ of
weight 3, 5 or 7  (i.e., there are no states in $\cF$ that are not
total derivatives). Indeed, we have not found a single case where the
number of conserved densities in $\cF$ for the Virasoro perturbation
$\phi_{1,3}$ exceeds those for generic $\beta$. 

The clarification of the duality breaking in
Toda theories and its repercussion in  the  perturbed conformal
models is certainly an interesting problem for future study.

\vspace{5mm}
\section{Acknowledgements}

GMTW was supported by an EPSRC (UK) advanced fellowship.
The work of PM was supported by NSERC (Canada).

GMTW would like to thank H.G. Kausch for several useful discussions on
section \ref{sec:ffn=0}, for providing ref. \cite{hgkunpub}, and for
pointing out reference \cite{BMP};
M. Freeman for discussions on \cite{DUA} and conserved densities in
general, and for a critical reading of the manuscript.

\vspace{10mm}

%-----------------------------------------------------
\newpage
\appendix

\section{Non-integrable supersymmetric 
evolution equations with an infinite
number of fermionic conservation laws}
\label{sec:fcq}

\def\dxdt{{\rm d}x{\rm d}\theta}

In this appendix, we present a neat example of a generic class of
equations that display an infinite number of fermionic conservation
laws but which are nonetheless non-integrable.\footnote{There is no
paradox in this statement: integrability of a field theory requires
not only an infinite number of conservation laws but also that these
conservation laws  be in involution.} 
We first  consider supersymmetric models.  In a second step, we show
that this curiosity is  not an artifact of supersymmetry, by
constructing a non-supersymmetric model with the same properties. 

Consider the general class of space supersymmetric evolution
equations: 
\be
  R_t = [M(R)]_{x} +aRR_x-\left[{(DR)(DM(R))\over R_x}\right]_x
\;,
\ee
where $a$ is a constant and $M(R)$ is an even arbitrary differential
polynomial in the superfield
$R$: 
\be
  R(x,\theta,t)= u(x,t) +\theta\sigma(x,t)
\;.
\ee
Here $u$ represents a commuting field, $\sigma$ is an 
anticommuting field, $\theta$ is an anticommuting space variable and
$D$ is the super-derivative: 
$D = \theta\partial_x +\partial_\theta, ~ D^2= \partial_x$.  $R$ is
thus an even superfield.  The special structure of the equation
readily implies the existence of an infinite number of conserved
integrals, which are of the form $\int \dxdt ~F(R)$ for any even
function $F(R)$ which can be written as  
\be
F(R) ={dG(R)\over dR} \equiv G(R)'
\;.
\ee 
Indeed, using the evolution equation, one has
\begin{eqnarray} 
    \int \dxdt ~F_t 
&=& \int \dxdt~\{F'[M(R)]_{x}+aF' RR_x + F''(DR)(D[M(R)])\} 
\nonumber \\
&=& \int \dxdt~\{F'[M(R)]_{x}-a FR_x - F'[M(R)]_{x}\} 
\nonumber\\
&=& -a \int \dxdt ~ G_x  ~ = ~0
\;.
\end{eqnarray}

However, these conservation laws are somewhat trivial. For instance,
with  
\be
  F(R)=R^n = u^n+\theta nu^{n-1}\sigma
\;,
\ee
they read
\be
  \int \dxdt~F(R) = \int \!{\mathrm d}x ~nu^{n-1}\sigma
\;,
\ee
and they disappear when $\sigma=0$.

The equation  we have considered is a
supersymmetric extension of 
\be
  u_t= [M(u)]_x+auu_x
\;,
\ee
which is generically non-integrable.  It is thus clear that for the
type of nonlinearity considered here, the supersymmetric process
itself is responsible for the infinite set of conservation
laws.\footnote{Actually, the same result follows with weaker
conditions, that is, without strict supersymmetry invariance -- see
below.} 

As a concrete example, consider the following supersymmetric extension
of the Korteweg-de Vries (KdV) equation: 
\be
  R_t = -R_{xxx} + 12 RR_x +\left[{(DR)(DR_{xx})\over R_x}\right]_x
\;,
\ee
whose component form reads
\begin{eqnarray} 
    u_t 
&=& -u_{xxx} +12uu_x +\left[{\sigma\sigma_{xx}\over u_x}\right]_x 
\;,\\
    \sigma_t 
&=& 12(\sigma u)_x 
   +\left[{\sigma \sigma_x\sigma_{xx} 
           -\sigma u_{xxx}u_x\over u_x^2}\right]_x
\;.
\end{eqnarray}
When $\sigma=0$, this reduces to the KdV equation. Notice that this
system differs from the two integrable fermionic extensions of the KdV
equation \cite{MRMK} (in which cases $\sigma$ has dimension $3/2$
while it is $5/2$ here). 

The infinite family of conservation laws found above does not
generalise the usual KdV conservation laws, whose leading term is of
the form
$\int {\rm d}x~(u^n+...)$.  
Conservation laws generalising the usual KdV ones involve non-local
expressions in $R$, with leading term $R^n(D^{-1}R)$ (the formalism
underlying the manipulations of such nonlocal charges is presented in 
\cite{DarM}).  It is a simple exercise to check that  
\be
  \int \dxdt~(D^{-1}R), \qquad\int \dxdt~R(D^{-1}R)
\;,
\ee
and  
\be
  \int \dxdt~[R^2(D^{-1}R)+{1\over 4}R_x(DR)]
\;,
\ee
are conserved. For KdV type equations, these conservation laws are not
remarkable: in a hydrodynamics context, they simply generalise the
conservation of mass, momentum and energy.  However, a nontrivial
conservation law of the form 
\be
\int \dxdt
{}~[R^3(D^{-1}R)+a_1R^2(DR_x)+a_2R_x^2(D^{-1}R)+a_3R_{xx}(DR_x)]
\;,
\ee
has not been found. For generalisations of the KdV equation, the
non-existence of such a conservation law is a clear signal of
non-integrability.\footnote{It is amusing to notice that a travelling
wave solution of the form $R=R(x-12ct,\theta, t)$ of the above
supersymmetric KdV equation  is related to the super Weierstrass
function.  Simple manipulations (i.e., integrate the resulting
equation once with respect to $x$, multiply the result by $R_x$, and
integrate again) yields 
\be
   R_x^2-4R^3-12cR^2-2(DR)(DR_x) +2k_1R+k_2 = 0
\;,
\ee
where $k_1, k_2$ are integration constants.  Now setting
$P = R+c ,\, g_2=2k_1+12c^2$ and $ g_3=k_2 -2k_1c-8c^3$,
one can rewrite the above equation as 
\be
  P_x^2 -4P^3+ g_2P + g_3 -2(DP)(DP_x) = 0
\;,
\ee
which is the defining equation for the super Weierstrass function
\cite{Freu}: $P = \wp(x;\tau) + \theta\delta
\partial_\tau\wp(x;\tau)$, where $\wp$ is the ordinary Weierstrass
function, $\tau$ is the modulus and $\delta$ is the super
modulus. $g_2(\tau)$ and $g_3(\tau)$ are the usual modular forms.}

With a simple deformation of the above supersymmetric KdV equation, we
can easily construct a family of non-supersymmetric systems having an
infinite number of fermionic conservation laws. One such deformation
is:  
\begin{eqnarray} 
    u_t 
&=& -a u_{xxx} +12uu_x +b\left[{\sigma\sigma_{xx}\over u_x}\right]_x 
\;,
\\
   \sigma_t 
&=& 12(\sigma u)_x 
    +\left[{b\,\sigma\sigma_x\sigma_{xx}-a\,\sigma u_{xxx}u_x\over
            u_x^2}\right]_x 
\;.
\end{eqnarray}
It is simple to check that for these equations the integrals  
$\int dx\, u^n\sigma $ are conserved for any value of $a$ and $b$.
However, the system is  invariant under a supersymmetric
transformation (defined by $\delta u = \epsilon \sigma$ and 
$\delta \sigma = \epsilon u_x$ with $\epsilon$ an anticommuting
parameter) only for $a=b$. 

We should stress that the fermionic character of the infinite sequence
of trivial conservation laws that have been constructed here is rooted
in the number of fermions (here one) in the system. If we consider
instead the coupling of two fermions with two bosonic fields, then we
could construct systems with an infinite number of {\it bosonic}
conservation laws quite easily.  Focusing on supersymmetric systems
for definiteness, it is simple to verify (along the above argument)
that for the following equation with $N=2$ supersymmetry (${\cal
R}=u+\theta_1 \sigma_1+\theta_2\sigma_2+\theta_1\theta_2 w$, $u,w$
bosonic and $\sigma_i$ fermionic and $D_i = \partial_{\theta_i}
+{\theta_i}\partial_x$)  
\[
  {\cal R}_t 
= -{\cal R}_{xxx} + 12 {\cal R}{\cal R}_x 
  + {1\over2}
    \left[{\sum_{i=1,2}D_i{\cal R}(D_i{\cal R}_{xx})\over{\cal R}_x}
            \right]_x
\]
the integrals $\int dx d\theta_1 d\theta_2 F({\cal R})$ -- which in
the present context are bosonic -- are conserved (with $F({\cal R}) =
dG/d{\cal R}$ for some $G$). 

\section{The operator product of two NS fields for $\mathbf{N=1}$.}

We are interested in the coefficients $\alpha$ and $\beta$ in the
operator products (with $\vec{k}\equiv \hphi_k(0)\vec{0}$):
\be
\begin{array}{rcl}
\hphi_{1}(1) \,\hG_{-1/2}\vec 2 &=& \hG_{-1/2}\vec{3} 
 + \left(\, \alpha \hL_{-1}\hG_{-1/2} + \beta \hG_{-3/2} \,\right) 
   \vec{3} + \ldots \;,\\
\end{array}
\label{eq.NNN}
\ee
where $\hphi_{i}$ is a Neveu-Schwarz field of weight $h_i$, and where
$\hphi_1$ is either fermionic $(\eta = 1)$ or bosonic $(\eta = -1)$.
Then, using the standard relations 
\bea
    (\hL_m - \hL_{m-1})\,\hphi_1(1) 
&=& \hphi_1(1)\,(\hL_m - \hL_{m-1} + h_1) 
\;,\\
    (\hG_m - \hG_n)\,\hphi_1(1) 
&=& -\eta\,\hphi_1(1)\,(\hG_m - \hG_n) 
\;,
\eea
we find that
\be
  \pmatrix{\alpha \cr \beta}
= \eta\,
  \pmatrix{2 h_3(2 h_3 + 1)   & 4 h_3 \cr
           4 h_3              & 2 h_3 + 2c/3 }^{-1} \,
  \pmatrix{ (h_2 + h_3 - h_1)(h_3 + h_1 - h_2) \cr 
             h_1 + h_2 - h_3 }
\;.\;\;
\ee
Inserting the values we use in section \ref{sec.n=1}, and denoting
$p'/p$ by $t$,
\be
  h_1 = \h_{r,1}  \;,\;\;
  h_2 = \h_{1,s}  \;,\;\;
  h_3 = \h_{r,s}  \;,\;\;
\ee
we find that in the two cases of interest, namely $(r,s) = (3,5)$ and
$(r,s) = (5,3)$, 
\be
\pmatrix{ \alpha & \beta }
=\left\{\begin{array}{ll}
  \left(\begin{array}{cc}
\frac{\eta \,t\,\left( -1 + 3\,t \right) }
     {3\,\left( 1 - 2\,t \right) \,{{\left( -1 + t \right) }^2}}   &
\frac{\eta \,\left( 5 - 15\,t + 12\,{t^2} \right) }
     {3\,{{\left( -1 + t \right) }^2}\,\left( -1 + 2\,t \right)}   
  \end{array}\right)
& (r,s) = (3,5)\;, \\
  \left(\begin{array}{cc}
\frac{\eta \,\left( 3 - t \right) \,t}
     {3\,\left( -2 + t \right) \,{{\left( -1 + t \right) }^2}}     &
\frac{\eta \,t\,\left( -12 + 15\,t - 5\,{t^2} \right) } 
     {3\,\left( -2 + t \right) \,{{\left( -1 + t \right) }^2}}     
  \end{array}\right)
& (r,s) = (5,3)\;.
  \end{array}\right.
\label{eq.abNNN}
\ee 

\section{The operator product of a R and a NS field for $\mathbf{N=1}$.}

We are interested in the coefficients $\alpha,\beta,A,B$ in the
operator products
\be
\begin{array}{rcl}
    \hphi_{NS}(1) \,\vec R 
&=& \vec{R'} 
    + \left(\, \alpha \hL_{-1} + \beta \hG_{-1} \,\right) \vec{R'} 
    + \ldots \\
    \hat \hG_{-1/2}\hphi_{NS}(1) \,\vec R 
&=& \vec{R'} 
    + \left(\, A \hL_{-1} + B \hG_{-1} \,\right) \vec{R'} + \ldots 
\end{array}
\label{eq.NRR}
\ee
where $\hphi_{NS}$ is a Neveu-Schwarz field of weight $H$, which is
either fermionic $(\eta = 1)$ or bosonic $(\eta = -1)$, 
$\vec R$ is a Ramond highest-weight state of $\hG_0$ eigenvalue
$\lambda$ and weight $h = \lambda^2 + c/24$, and $\vec{R'}$ is a
Ramond highest-weight state of $\hG_0$ eigenvalue $\lambda'$ and
weight $h' = \lambda'^2 + c/24$.  Then, using the standard relations: 
\bea
  (\hL_1 - \hL_0)\,\hphi_{NS}(1) &=& \hphi_{NS}(1)\,(\hL_1 - \hL_0 + H) 
\;,\\
  (\hG_1 - \hG_0)\,\hphi_{NS}(1)&=&-\eta\,\hphi_{NS}(1)\,(\hG_1-\hG_0)
\;,\\
  \hG_0        \,\hphi_{NS}(1) &=&  \hG_{-1/2}\hphi_{NS}(1) \;
                              -\eta \,\hphi_{NS}(1) \hG_0
\;,
\eea
we find that
\be
  \pmatrix{A \cr B}
= \pmatrix{\lambda' + \eta \lambda & 2 \cr
           1/2                     & -\lambda' + \eta \lambda}
  \pmatrix{\alpha \cr \beta}
\;,\;\;
  \pmatrix{\alpha \cr \beta}
= \pmatrix{ 2h'         & 3\lambda'/2 \cr 
            3\lambda'/2 & 2h'+c/4        }^{-1}
  \pmatrix{H + h' - h \cr \lambda' + \eta \lambda}
\;.\;\;
\ee
Inserting the values we use in section \ref{sec.n=1}, and denoting
$p'/p$ by $t$,
\be
  H = \h_{1,s} = \frac{ (1 - s t)^2 - (1-t)^2 }{ 8t } \;,\;\;
  \lambda = \lambda_{r,1} = \frac{ r - t }{\sqrt{8t}} \;,\;\;
  \lambda'= -\eta\lambda_{r,s} = -\eta\frac{ r - st}{\sqrt{8t}}
  \;,\;\;
\ee
we find that
\bea 
 \beta &=&
 \eta  \frac{\left( -1 + s \right) \,{t^{3/2}}}
            { \sqrt 2 \,\left( -2 + r + t - s\,t \right) \,
                        \left( 1 - r - 2\,t + s\,t \right) }
\;,
\label{eq.beta}
\\ 
  B &=& 
   \frac{\left( -1 + s \right) \,t\,
       \left( 4 - 4\,r - 3\,t + 3\,s\,t \right) }
     {4\,\left( 1 - r - 2\,t + s\,t \right) \,
       \left( 2 - r - t + s\,t \right) }
\;.
\label{eq.B}
\eea 

\newpage
\section{Solutions to $\mathbf{\Delta{\hsv}=3/2,2,5/2}$}
In these tables we take $(p-s)(r-1) \geq (p'-r)(s-1)$. The cases
$(p-s)(r-1)<(p'-r)(s-1)$ may be obtained by taking $p'\leftrightarrow
p$ and $r \leftrightarrow s$.

\begin{table}[!htb]
{\scriptsize
\renewcommand{\arraystretch}{1.4}
\[
\begin{array}{|cccccl|cc|cc|c|}
\hline
(p-s)(r-1) &+& (p'-r)(s-1) &+& (r-1)(s-1) &=3 
&(r,s)&(p,p')&NS&R&\hbox{Min.}\\
\hline
3 && 0 && 0 && \left\{ \matrix{ (2,1) \cr (4,1) }\right.
             &         \matrix{ (4,-) \cr (2,-) }        
             &         \matrix{       \cr \cr  }        
             &         \matrix{ \tik  \cr \tik  }        
             &         \matrix{ \tik  \cr \tik  }        \\

2 && 1 && 0 && \multicolumn{2}{c|}{\hbox{ No solutions }}
             &&& \\
2 && 0 && 1 && (2,2) & (2,4) &&& \\

1 && 1 && 1 && (2,2) & (3,3) &&& \\
1 && 0 && 2 && (2,3) & (4,2) &&& \\

0 && 0 && 3 && \left\{ \matrix{ (2,4) \cr (4,2) }\right.
             &         \matrix{ (4,2) \cr (2,4) }        
             &&&    \\
\hline
\hline
(p-s)(r-1) &+& (p'-r)(s-1) &+& (r-1)(s-1) &=4 
 &(r,s)&(p,p')&NS&R&\hbox{Min.}\\
\hline
4 && 0 && 0 && \left\{ \matrix{ (2,1) \cr (3,1) \cr (5,1) }\right.
             &         \matrix{ (5,-) \cr (3,-) \cr (2,-) }        
             &         \matrix{       \cr \tik  \cr \tik  }        
             &         \matrix{ \tik  \cr       \cr \cr   }        
             &         \matrix{ \tik  \cr \tik  \cr \tik  }   \\

3 && 1 && 0 && \multicolumn{2}{c|}{\hbox{ No solutions }}
             &&& \\
3 && 0 && 1 && (2,2) & (5,2) &&& \\

2 && 2 && 0 && \multicolumn{2}{c|}{\hbox{ No solutions }}
             &&& \\
2 && 1 && 1 && (2,2) & (4,3) &&& \\
2 && 0 && 2 && \left\{ \matrix{ (3,2) \cr (2,3) }\right.
             &         \matrix{ (3,3) \cr (5,2) }        
             &&& \\

1 && 1 && 2 && \multicolumn{2}{c|}{\hbox{ No solutions }}
             &&& \\
1 && 0 && 3 && (2,4) & (5,2) &&& \\

0 && 0 && 4 && \left\{ \matrix{ (5,2) \cr (3,3) \cr (2,5) }\right.
             &         \matrix{ (2,5) \cr (3,3) \cr (5,2) }        
             &&& \\
\hline
\hline
(p-s)(r-1) &+& (p'-r)(s-1) &+& (r-1)(s-1) &=5
 &(r,s)&(p,p')&NS&R&\hbox{Min.}\\
\hline
5 && 0 && 0 && \left\{ \matrix{ (2,1) \cr (6,1) }\right.
             &         \matrix{ (6,-) \cr (2,-) }        
             &         \matrix{       \cr \cr   }        
             &         \matrix{ \tik  \cr \tik  }        
             &         \matrix{ \tik  \cr \tik  }        \\
4 && 1 && 0 && \multicolumn{2}{c|}{\hbox{ No solutions }}
             &&& \\
4 && 0 && 1 && (2,2) & (5,2) &&& \\

3 && 2 && 0 && \multicolumn{2}{c|}{\hbox{ No solutions }}
             &&& \\
3 && 1 && 1 && (2,2) & (5,3) &\ttk&&\tik \\
3 && 0 && 2 && (2,3) & (6,2) &&& \\

2 && 2 && 1 && (2,2) & (4,4) &&& \\
2 && 1 && 2 && (2,3) & (3,4) &&& \\
2 && 0 && 3 && (2,4) & (6,2) &&& \\

1 && 1 && 3 && \multicolumn{2}{c|}{\hbox{ No solutions }}
             &&& \\
1 && 0 && 4 && (2,5) & (6,2) &&& \\

0 && 0 && 5 &&\left\{ \matrix{ (2,6) \cr (6,2) }\right.
             &         \matrix{ (6,2) \cr (2,6) }        
             &&& \\
\hline

\end{array}
\]
\caption{Solutions to $\Delta{\hsv}=3/2,2,5/2$}
\label{tab:2b}
\label{tab:2c}
\label{tab:2d}
}
\end{table}

%---------------------------------------------------------


\newpage
\begin{thebibliography}{1}
{
\small
\parskip 0pt

\bibitem{ONE}
P.~Mathieu and G.M.T.~Watts, 
\newblock 
{\it Probing integrable perturbations of conformal theories using
singular vectors},  
Nucl.\ Phys.\ {\bf B475} (1996) 361--396, hep-th/960308.


\bibitem{ZAMO}
A.B.~Zamolodchikov,
\newblock JETP Lett. {\bf 46} (1987) 160;
\newblock Int. Jour. Mod. Phys. {\bf A3} (1988) 4235;
\newblock Adv.\ Stud.\ in Pure Math.. 19 (1989) 614.

\bibitem{DUA}
B.L.~Feigin and E.~Frenkel,
\newblock {\em Integrals of motion and quantum groups},
\newblock proceedings of C.I.M.E. Summer School on `Integrable systems
and Quantum groups', 1993
\newblock {\tt hep-th/9310022} (revised March 95).
%H.~G.~Kausch and G.~M.~T.~Watts,
%\newblock Nucl.~Phys.~{\bf B386} (1992) 166.


\bibitem{FKM}
P.G.O.~Freund, T.R.~Klassen and E.~Melzer,
\newblock Phys. Lett. {\bf B229} (1989) 243.


\bibitem{EY}
T.~Eguchi and S.K.~Yang,
\newblock Phys.~Lett.~{\bf B235} (1990) 282;
\newline
A.~Kuniba,  T.~Nakanishi and J.~Suzuki,
\newblock Nucl.~Phys.~{\bf B356} (1991) 750.

\bibitem{SV}
P.~Di~Francesco and P. Mathieu,
\newblock Phys.~Lett.~{\bf B278}  (1992) 79.

\bibitem{Ma}
P.~Mathieu,
\newblock Nucl.~Phys.~{\bf B336} (1990) 338; 
K. Schoutens, Nucl.~Phys.~{\bf B344} (1990) 665 

\bibitem{DeMa}
D.~Depireux and P.~Mathieu, Phys. Lett. {\bf B 308} (1993) 2729. 

\bibitem{Melzer}
E.~Melzer,  {\it Supersymmetric analogs of the Gordon-Andrews
identities, and related TBA systems}, {\tt hep-th/9412154}. 

\bibitem{Vays}
I.~Vaysburd,  Nucl.Phys.{\bf B446} (1995) 387, {\tt hep-th/9503070}.

\bibitem{DGZ}
{G.W. Delius, M.T. Grisaru and D. Zanon},
 % {\it Exact S-Matrices for non-simply-laced affine Toda theories}, 
Nucl. Phys. {\bf B382} (1992) {365--408}, hep-th/9201067.

\bibitem{FF}
B.L. Feigin and D.B Fuchs,
\newblock {\em Representations of the Virasoro algebra},
\newblock in `{Representations of infinite-dimensional Lie
groups and Lie algebras}', eds.\ A.\ Vershik and D.\ Zhelobenko, 
Gordon and Breach (1989)

\bibitem{NU}
A. Kent, PhD thesis, Cambridge University (1986); 
\newline
V. Kac and M. Wakimoto,
{Proc. Nat. Acad. Sci. USA}, {\bf 85} (1988) 4956; 
\newline
P. Mathieu, D. S\'en\'echal and M. Walton, Int. J. Mod Phys {\bf A7}
Suppl. 1B, 731 (1992), {\it Proceedings of the RIMS Research Project
1991 Infinite Analysis}, Adv. Series in Math. Phys., vol 16 (1992).

\bibitem{DarM}
P. Dargis and P. Mathieu,  
Phys. Lett. {\bf A176} (1993) 67.

\bibitem{olsh1}
M.A. Olshanetsky, 
  %  {\it Supersymmetric Two-Dimensional Toda Lattice},
Commun. Math. Phys. {\bf 88} (1983) 63--76.

\bibitem{LMan1}
H.C. Liao and P. Mansfield,
  %  {\it Light-cone quantization of Super-Liouville Theory},
Nucl. Phys. {\bf B344} (1990) {696-730}

\bibitem{DF}
Vl.S. Dotsenko and V. Fateev, 
  %  \newblock {\it Conformal algebra and multiploint correlation
  %  functions in 2D statistical models}, 
\newblock Nucl. Phys. {\bf B240 } (1984) 312;
 % \newblock {\it Four-Point correlation functions and  the operator
 % algebra in 2D conformal invariant theories},  
\newblock Nucl. Phys. {\bf B251 } (1985) 691.

\bibitem{Feld1}
G. Felder,
  %  \newblock {\it BRST Approach to Minimal Models},
\newblock Nucl. Phys. {\bf B317} (1989) 215.

\bibitem{BMP}
P. Bouwknegt, J. McCarthy and K. Pilch,
  %  \newblock {\em Fock space resolutions of the Virasoro
  %  highest-weight modules with $c\leq 1$},
\newblock Lett. Math. Phys. 23 (1991) 193, 
\newblock hep-th/9108023

\bibitem{svir}
M.A. Bershadsky, V.G. Knizhnik and M.G. Teitelman,
  %  \newblock{\it Superconformal symmetry in two dimensions},
\newblock Phys. Lett. B151 (1985) 31;
\newline
D. Friedan, Z. Qiu and S. Shenker,
 % \newblock{\it Superconformal invariance in two dimensions and the
 % tricritical Ising model}, 
\newblock Phys. Lett. B151 (1985) 37.

\bibitem{FL}
S.L.~Lukyanov and V.A.~Fateev
Sov. Sci. Rev. A Phys. {\bf 15} (1990) 1.

\bibitem{Wat}
G.M.T.~Watts, Nucl. Phys. {\bf B339} (1990) 177. 

\bibitem{fermcls}
L. Palla,
%{\it Perturbed $W$ algebras and affine Toda theories},
\newblock 
Nucl. Phys. {\bf B341} (1990) 714--741;

\bibitem{hgkunpub}
H.G. Kausch, {\it Notes on Fock modules}, unpublished.

\bibitem{MRMK}
Yu.I. Manin and A. O. Radul, Comm. Math. Phys. {\bf 114} (1985) 65;
\newline
P. Mathieu, J. Math. Phys. {\bf 29} (1988) 2499; 
\newline
B. A. Kupershmidt, Phys. Lett. A {\bf 102} (1984) 213.

\bibitem{Freu}
J.M. Rabin and P.G.O. Freund, Comm. Math. Phys. {\bf 114} (1988) 131.

}

\end{thebibliography}
\end{document}